\documentclass[twocolumn,aps,pr,superscriptaddress,preprintnumbers,nofootinbib,10pt]{revtex4-2}
\usepackage{amsmath,amssymb}
\usepackage[dvipdf,dvips]{graphicx}
\usepackage{color}
\usepackage{hyperref}
\usepackage{url}
\usepackage{slashed}
\usepackage{subfigure}
\usepackage[usenames,dvipsnames]{xcolor}
\usepackage{amsmath}
\usepackage{amsfonts}
\usepackage{float} 
\usepackage{amssymb}
\usepackage{epsfig}
\usepackage{graphics}
\usepackage{euscript}
\usepackage{slashed}
\usepackage{epstopdf}
\usepackage[utf8]{inputenc}
\allowdisplaybreaks
\usepackage[normalem]{ulem}
\usepackage{pifont}
\usepackage{dsfont}
\usepackage{MnSymbol}
\usepackage{verbatim}
\usepackage{graphicx}
\usepackage{latexsym}

\hypersetup{
colorlinks=true,
citecolor=blue,
citebordercolor=red,
linktoc=all,
linkcolor=blue,
urlcolor=blue
}

\def \s{\mathsf{s}}

\def \and{\textmd{and}}

\newbox{\ORCIDicon}
\sbox{\ORCIDicon}{\large
                  \includegraphics[width=0.8em]{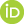}}

\begin{document}

\title{Perturbative approach to the infrared gluon propagator in the maximal Abelian gauge}

\author{D.~M.~van~Egmond,\href{https://orcid.org/0000-0003-0019-7651}{\usebox{\ORCIDicon}}}\email{duifjemaria@gmail.com}
\affiliation{ICTP South American Institute for Fundamental Research Instituto de F\'isica Te\'orica,
UNESP - Univ. Estadual Paulista, Rua Dr. Bento Teobaldo Ferraz, 271, 01140-070 S\~ao Paulo, SP, Brazil}
\author{L.~C.~Ferreira} \email{luigicferreira@gmail.com}
\affiliation{EEEFM Agenor de Souza Lé, S/N, R. Alan Kardec - Divino Espírito Santo, Vila Velha - ES, 29107-240}
\author{A.~D.~Pereira\,\href{https://orcid.org/0000-0002-6952-2961}{\usebox{\ORCIDicon}}} \email{adpjunior@id.uff.br}
\affiliation{Instituto de F\'isica, Universidade Federal Fluminense, Campus da Praia Vermelha, Av. Litor\^anea s/n, 24210-346, Niter\'oi, RJ, Brazil}
\author{G.~Peruzzo,\href{https://orcid.org/0000-0002-4631-5257}{\usebox{\ORCIDicon}}}\email{gperuzzofisica@gmail.com}
\affiliation{Instituto de F\'isica, Universidade Federal Fluminense, Campus da Praia Vermelha, Av. Litor\^anea s/n, 24210-346, Niter\'oi, RJ, Brazil}
\author{S.~P.~Sorella} \email{silvio.sorella@gmail.com}
\affiliation{Universidade do Estado do Rio de Janeiro, Instituto de F\'isica –Departamento de F\'isica Te\'orica – Rua S\~ao Francisco Xavier 524, 20550-013, Maracan\~a, Rio de Janeiro, Brazil}

%%%%%%%%%%%%%%%%%%%%%%%%%%%%%%%%%%%%%%%%%%
\begin{abstract}
 The inclusion of a mass-like term for the gluon in Yang-Mills theories quantized in the Landau gauge has proven to be an effective way of reproducing lattice results for gauge-fixed correlation functions within perturbative computations. Since those quantities are gauge dependent, it is natural to question how general this prescription is for describing the infrared behavior of gluon and Faddeev-Popov ghost propagators in different gauges. In this work, we provide a systematic investigation of this issue in the maximal Abelian gauge, which cannot be deformed into the Landau gauge and has been investigated in gauge-fixed lattice simulations. We compute the one-loop non-Abelian and diagonal gluon propagators and perform fits to lattice data in the case of $SU(2)$. Our results show that the transverse component of the non-Abelian gluon propagator as well as the diagonal gluon propagator, are in good agreement with lattice data in the infrared.
\end{abstract}
%%%%%%%%%%%%%%%%%%%%%%%%%%%%%%%%%%%%%%%%%%

\maketitle

%-------------------------------------------------------
\section{Introduction \label{Sec:Intro}}
%-------------------------------------------------------

Accessing the infrared (IR) regime of Yang-Mills (YM) theories in the continuum remains a challenging task. Since these theories are strongly coupled at low energies, non-perturbative methods are required. Concrete calculations in the continuum require fixing the gauge which is usually done through the Faddeev-Popov (FP) procedure  \cite{Faddeev:1967fc}. For this reason, many explicit computations are focused on gauge-dependent correlators that in turn build up the gauge-invariant correlation functions associated with observables. In recent years, important progress has been made in understanding gauge-fixed correlators through numerical simulations \cite{Sternbeck:2007ug,Cucchieri:2007md, Cucchieri:2007rg,Bornyakov:2008yx,Bogolubsky:2009dc,Dudal:2010tf,Maas:2011se,Cucchieri:2011ig,Oliveira:2012eh,Dudal:2018cli} and (semi-)analytical approaches \cite{Brambilla:2014jmp, Fischer:2008uz, Aguilar:2008xm,Dudal:2007cw, Dudal:2008sp,Pelaez:2021tpq, Pelaez:2017bhh} approaches with the Landau gauge being employed in most studies. Gauge-fixed lattice simulations offer, among other things, an independent benchmark for the computations performed in the continuum and such synergetic approach has led to considerable progress in the comprehension of the infrared behavior of correlation functions in Yang-Mills theories over the past decades. 

It is well known that gauge fixing in non-Abelian gauge theories suffers from a subtle issue in the non-perturbative regime due to the presence of Gribov copies \cite{Gribov:1977wm,Singer:1978dk}, see, e.g., \cite{Sobreiro:2005ec,Vandersickel:2011zc,Vandersickel:2012tz} for reviews on the Gribov problem. The FP procedure is not able to fix completely the gauge redundancy and spurious configurations, the Gribov copies, remain present in the path integral. It is still unkown how to completely eliminate those configurations and only a partial resolution of the issue is understood for some particular gauge choices. In fact, most of the works in this direction are restricted to the Landau gauge \cite{Gribov:1977wm,Zwanziger:1989mf,Dudal:2008sp} where a systematic elimination of infinitesimal Gribov copies is achieved. Another gauge in which the Gribov problem has been partially dealt with is the so-called maximal Abelian gauge (MAG), see, e.g., \cite{Capri:2005tj,Gongyo:2013rua}. It treats non-Abelian (off-diagonal) and Abelian (diagonal) components of the gauge field differently. This feature makes the MAG well-suited to study color confinement via the dual superconductivity mechanism \cite{Nambu1974,tHooft1975}, where low-energy Yang-Mills theories reduce to an effective Abelian theory with monopoles \cite{Mandelstam1976,Ezawa1982,Suzuki1990,Suzuki:1992gz,Hioki1991,Sakumichi:2014xpa}.

For all gauges in which a concrete (partial) elimination of Gribov copies was implemented, the analysis was restricted to infinitesimal copies. Those are configurations that are related by an infinitesimal gauge transformation and satisfy the same gauge condition. The action that implements the elimination of the infinitesimal copies is the so-called Gribov-Zwanziger (GZ) action, see \cite{Zwanziger:1989mf}. Originally, it was constructed in the Landau gauge leading to a local and renormalizable framework which effectively takes into account the elimination of infinitesimal Gribov copies. It was found that the GZ action features IR instabilities that are estabilized by the inclusion of condensates. The resulting theory is the so-called Refined Gribov-Zwanziger (RGZ) action, see \cite{Dudal:2008sp}. Later, the construction of the analogue of the GZ action in the MAG was achieved in \cite{Capri:2005tj}. The dynamical formation of condensates, whose inclusion led to the so-called RGZ action was worked out in \cite{Capri:2008ak,Capri:2008vk}. Remarkably, the RGZ action (both in the Landau gauge and in the MAG) is perturbatively renormalizable and local. More recently, the construction of the RGZ action in linear covariant and Curci-Ferrari gauges was achieved, see \cite{Capri:2015ixa,Capri:2015nzw,Capri:2016aif,Capri:2016ovw,Capri:2017bfd,Pereira:2016fpn}. It is worth mentioning that different strategies to deal with Gribov copies were employed in \cite{Serreau:2012cg,Serreau:2013ila,Serreau:2015yna,Reinosa:2020skx} where instead of eliminating the spurious configurations, an averaging procedure is used. Recently, attempts to establish a connection between both approaches were carried out in \cite{CarmoTerin:2025agt,CarmoTerin:2026rzb}. The RGZ action can be viewed as an improved version of the gauge-fixed FP action since it partially deals with the presence of Gribov copies. Nevertheless, the removal of infinitesimal Gribov copies leads to an effective modification of the tree-level gluon propagator as well as new interactions with auxiliary localizing fields. As a result, already at tree-level the gluon propagator carries non-perturbative information and its form is such that it fits very well with the lattice data in the infrared. This is well explored in the Landau gauge \cite{Dudal:2007cw,Cucchieri:2011ig,Oliveira:2012eh,Dudal:2018cli} and reasonably understood in the MAG \cite{Capri:2008ak}.

Recent lattice simulations have revealed a massive-like behavior for the gluon propagator in the deep IR \cite{Sternbeck:2007ug,Cucchieri:2007md, Cucchieri:2007rg,Bornyakov:2008yx,Bogolubsky:2009dc}. This has also been seen in continuum approaches as in the RGZ,  and in functional methods, see, e.g., \cite{Aguilar:2008xm,Fischer:2008uz,Eichmann:2021zuv,Horak:2022aqx,Ferreira:2025tzo} and in the context of center vortices \cite{Junior:2025gxg}. From the RGZ perspective, for instance, the tree-level gluon propagator can be written as a Yukawa-like propagator with a momentum-dependent mass. This feature led \cite{Tissier:2010ts,Tissier:2011ey} to propose an effective massive model for the gluon, the Curci-Ferrari (CF) model, to describe the IR behavior of the gluon propagator obtained in lattice simulations in the Landau gauge. First, by including one-loop corrections, they were able to fit the gluon and ghost propagators in the IR regime. Later, these results were improved with the inclusion of two-loop corrections \cite{Gracey:2019xom}. Alongside this model, a renormalization scheme was proposed, the infrared-safe scheme, in which the effective coupling does not develop an IR Landau pole. We refer to \cite{Pelaez:2021tpq} for a review and to \cite{Reinosa:2024vph} for a pedagogical introduction to the CF model. More developments, including dynamical quarks \cite{Pelaez:2014mxa}, such as the computation of the (un)quenched ghost-gluon and three-gluon vertices \cite{Pelaez:2013cpa,Figueroa:2021sjm}, quark-gluon vertex  \cite{Pelaez:2015tba,Barrios:2020ubx}, and the four-point gluon vertex \cite{Barrios:2024ixj} were carried out within the CF model. The inclusion of quarks yields more realistic investigations, which can be related to hadronic phenomenology, see \cite{Bopsin:2025vhz,Alvez:2025wek}. At finite temperature, the model can be employed to probe the breaking of center symmetry which is associated with the confinement/deconfinement transition of quarks and gluons, see \cite{Reinosa:2014ooa, vanEgmond:2021jyx, vanEgmond:2025zxf}. At this stage, a comment is in order: The CF model consists in introducing a mass-like term on top of the gauge-fixing term (originally, in the Landau gauge). It should not be thought as a massive Proca-like theory. Effectively, the introduction of such a mass-like term can be thought as the most economical model to deal with the existence of (infinitesimal) Gribov copies 

After gauge-fixing through the FP procedure, the gauge(-parameter) independence of physical quantities is controlled by the BRST-symmetry \cite{Becchi:1975nq,Tyutin:1975qk}. It allows for a systematic definition a physical subspace through its associated charge. However, beyond the perturbative regime this construction can be challenged due to the existence of Gribov copies\footnote{In fact, as originally formulated, the (R)GZ action breaks BRST symmetry explicitly but in a soft manner. More recently, a BRST-invariant formulation of the RGZ action in linear covariant gauges was proposed, see, e.g., \cite{Capri:2015ixa,Capri:2017bfd} thanks to the use of dressed variables.}. The CF model contains the main elements of the BRST-quantization, namely, the gauge-fixing term with the correspond FP ghost fields. In fact, the mass-like term for the gluons is added together with the gauge-fixed YM action in the Landau gauge. However, the mass-like term explicitly breaks BRST symmetry.  In this model there is a symmetry which resembles the BRST, but it is not nilpotent. This later property is crucial to guarantee that ghost states with negative norm decouples \cite{Kugo:1979gm}. This poses a limitation to the CF model. For instance, if one adds a simple mass-like term for the gluons in the gauge-fixed YM action in linear covariant gauges, the lack of BRST invariance will translate into gauge-parameter dependence of gauge-invariant correlators quantities. Progress in building a CF-like model in linear covariant gauges can be found in \cite{Comitini:2023urc}. The construction of a CF-like model in the Curci-Ferrari-Delbourgo-Jarvis gauge has also been worked out in \cite{Serreau:2015yna} and \cite{Cabrera:2026arc} with quarks. 

In light of these considerations, we believe it interesting to investigate whether CF-like models in other gauges that are not simply related to the Landau gauge by a suitable choice of gauge parameter are able to reproduce lattice data as effectively as in the Landau gauge. The main reason behind this is simplicity, i.e., the possibility of having an effective model that access the IR dynamics of YM theories through explicit perturbative computations. Therefore, we propose an analogous CF-like model in the MAG with gauge group $SU(2)$. This gauge brings new challenges with respect to the Landau gauge due to its non-linearity. In particular, this engenders the introduction of ghost quartic interactions and a fiducial parameter $\alpha$ to be renormalizable. Only after explicit computations are performed one takes the limit $\alpha \to 0$ and the MAG is recovered. Moreover, it breaks the global color symmetry explicitly. Interestingly, however, this gauge can be implemented in the lattice, as was done in the four-dimensional case for $SU(2)$ in \cite{Amemiya:1998jz,Bornyakov:2002vv,Bornyakov:2003ee} and has also been explored by functional methods, see \cite{Huber:2010tvj}. More recent simulations for $SU(3)$ were also reported in \cite{Gongyo:2012jb}. Unfortunately, moving from $SU(2)$ to $SU(3)$ in the case of the MAG is not so simple due to the existence of extra vertices and we focus on the $SU(2)$ case in this work for simplicity.

The effective model that we propose for the $SU(2)$ MAG contains the same elements of the CF model. We keep the gauge-fixing term of the renormalizable MAG, i.e., we include the quartic ghost terms with the gauge parameter $\alpha$, and add mass-like terms for the gauge fields. These masses should play the crucial role to capture the behavior of the gluon propagators for $p\rightarrow 0$ seen on the lattice. Since the global color symmetry is explicitly broken in this gauge, we consider different masses for the components, namely, $M$ and $m$ for the off-diagonal and diagonal components, respectively. In fact, to properly describe the Abelian dominance \cite{Bornyakov:2003ee} observed in the IR region, we need $M^2 > m^2$. We also add a mass term for the off-diagonal ghosts, which is necessary to have a renormalizable model as will be clear later. Since on the lattice the original MAG ($\alpha=0$) can be simulated, in the end of any calculation with this model the limit $\alpha\rightarrow 0$ should be taken for meaningful comparisons.

This article is organized as follows. In Section \ref{Sec:Model}, we present and motivate the IR-effective model for the $SU(2)$ MAG. In Section \ref{Sec:GP1L}, we present the one-loop results for the off-diagonal gluon propagator and the diagonal gluon propagator, alongside with a comparison with the lattice results. Finally, in Section \ref{Sec:Conclusions}, we collect our conclusions and indicate some directions for future research.

%-------------------------------------------------------
\section{The Model \label{Sec:Model}}
%-------------------------------------------------------

%-------------------------------------------------------
\subsection{Yang-Mills theories in the MAG \label{Sec:Model.1}}
%-------------------------------------------------------

The model we consider in this work corresponds to Euclidean four-dimensional Yang-Mills theories in the MAG with $SU(2)$ gauge group equiped with mass-like terms for the Abelian and non-Abelian components of the gauge field. Before discussing the introduction of the mass terms, we start with a short overview of the quantization of YM theories in the MAG. The starting point is the YM action given by\footnote{We employ the short-hand notation $\int {\rm d}^4x \equiv \int_x$.}
\begin{eqnarray}
S_{\mathrm{YM}} &=&\frac{1}{4}\int_x~F^{A}_{\mu\nu}F^{A}_{\mu\nu} \nonumber\\
&=&  \frac{1}{4}\int_x\left(F^{a}_{\mu\nu}F^{a}_{\mu\nu} + F_{\mu\nu}F_{\mu\nu}\right)\,,
\label{model1}
\end{eqnarray}
where $F^{a}_{\mu\nu} = \EuScript{D}^{ab}_\mu A^{b}_{\nu}-\EuScript{D}^{ab}_\nu A^{b}_{\mu}$ and $F_{\mu\nu} = \partial_\mu A_\nu - \partial_\nu A_\mu +g\epsilon^{ab}A^{a}_{\mu}A^{b}_{\nu}$. The generators of the gauge group are $\left\{T^A\right\}$ with $A=1,2,3$, the Cartan subgroup is generated by $T^3$ and therefore we employ small latin indices $a,b,\ldots = 1,2$ to represent the non-Abelian components of the algebra. Since we just have one Abelian component, we omit its associated index, i.e., $A^3_\mu \equiv A_\mu$. The covariant derivative $\EuScript{D}^{ab}_\mu$ is defined only with respect to the Abelian component, i.e., $\EuScript{D}^{ab}_\mu \equiv \delta^{ab}\partial_\mu - g\epsilon^{ab}A_\mu$. Finally, the anti-symmetric symbol $\epsilon^{ab}$ is defined by $\left[T^A,T^B\right] = {i\epsilon ^{ABC}T^C} $ where $\epsilon^{ABC}$ is  the Levi-Civita symbol and $\epsilon^{ab3} \equiv \epsilon^{ab}$.

The BRST quantization of YM theories in the MAG requires the introduction of the following BRST-exact term,
\begin{eqnarray}
S_{\mathrm{FP}}&=&\s\int_x\bar{c}^a \left(\EuScript{D}^{ab}_{\mu}A^{b}_{\mu}-\frac{\alpha}{2}ib^a+g\frac{\alpha}{2}\epsilon^{ab}\bar{c}^b c\right)\nonumber\\
&+&\s\int_x\bar{c}\,\partial_\mu A_\mu\,.
\label{model2}
\end{eqnarray}
The nilpotent BRST operator $\s$ acts on the elementary fields as
\begin{align}
\s A^a_\mu &= -\EuScript{D}^{ab}_{\mu}c^b - g\epsilon^{ab}A^b_\mu c\,, &&\s A_\mu = -\partial_\mu c - g\epsilon^{ab}A^a_\mu c^b\,,\nonumber\\
\s c^a &= g\epsilon^{ab}c^b c\,, &&\s c = \frac{g}{2}\epsilon^{ab}c^a c^b\,,\nonumber\\
\s\bar{c}^a &= i b^a\,, &&\s\bar{c} = ib\,,\nonumber\\
\s b^a &= 0\,, &&\s b= 0\,,
\label{model3}
\end{align}
where $(\bar{c}^a,c^a)$ and $(\bar{c},c)$ are the non-Abelian and Abelian FP ghosts. The fields $b^a$ and $b$ are the non-Abelian and Abelian Nakanishi-Lautrup fields, respectively. The resulting FP action is 
\begin{eqnarray}
S_{\rm FP} &=& \int_x\,\Bigg[ib^a\EuScript{D}^{ab}_{\mu}A^{b}_{\mu}+\frac{\alpha}{2}b^a b^a+ib\,\partial_\mu A_\mu-\bar{c}^{a}\mathcal{M}^{ab}c^b\nonumber\\
&+&g\epsilon^{ab}\bar{c}^a\left(\EuScript{D}^{bc}_{\mu}A^{c}_{\mu}\right)c+\bar{c}\,\partial_\mu(\partial_\mu c + g\epsilon^{ab}A^{a}_{\mu}c^b)\nonumber\\
&+&\alpha g \epsilon^{ab}ib^a \bar{c}^b c -\frac{\alpha}{2}g^2(\bar{c}^a c^a)^2\Bigg]\,,
\label{model4}
\end{eqnarray}
with $\mathcal{M}^{ab}\equiv -\EuScript{D}^{ac}_{\mu}\EuScript{D}^{cb}_{\mu}-g^2\epsilon^{ac}\epsilon^{bd}A^{c}_{\mu}A^{d}_{\mu}$ being the Faddeev-Popov operator in the MAG and $\alpha$ is a gauge parameter. Some comments are in order. The ``true" MAG is attained at vanishing value for the gauge parameter, i.e., $\alpha=0$. This means, that the MAG selects gauge field configurations which satisfy
\begin{equation}
\EuScript{D}^{ab}_\mu A^b_\mu = 0\,,
\label{model5}
\end{equation}
and the residual $U(1)$ symmetry is fixed by a Landau-type condition, i.e., $\partial_\mu A_\mu = 0$. The condition \eqref{model5} is achieved by demanding the minimization of the functional 
\begin{equation}
\mathcal{R} [A] = \int_x A^a_\mu A^a_\mu\,,
\label{model6}
\end{equation}
along each gauge orbit, which makes the search for gauge fields satisfying \eqref{model5} algorithmic for gauge-fixed lattice simulations. Yet condition \eqref{model5} is non-linear and requires the introduction of quartic interactions in the FP ghosts and thereby we define the FP action containing $\alpha$-dependent terms that are necessary for renormalizability \cite{Kondo:2000zva,Fazio:2001rm,Shinohara:2001cw}. The non-linear structure of the MAG also leaves imprints in the formation of condensates. In particular, the off-diagonal components of ghosts and gluons acquire a dynamical mass, see \cite{Kondo:2001nq,Dudal:2002xe,Dudal:2004rx}. This leads to a hint towards the Abelian dominance phenomenon which nicely fits the picture of non-Abelian degrees of freedom decoupling below the dynamical-mass scale and rendering an effective Abelian theory in the deep infrared, see, e.g., \cite{tHooft:1981bkw,Ezawa:1982bf,Suzuki:1989gp,Kondo:1997pc}. The analytic treatment of the formation of condensates was achieved by the introduction of a BRST-on-shell invariant dimension two ghost-gluon operator \cite{Kondo:2001nq,Dudal:2004rx}, i.e.,
\begin{equation}
\EuScript{O}(\alpha) = \frac{1}{2}A^a_\mu A^a_\mu - \alpha \,\bar{c}^a c^a\,.
\label{model7}
\end{equation}
This operator can be included in a BRST-invariant manner by means of the so-called Local Composite Operator (LCO) technique as follows: two external sources $\lambda$ and $J$ are introduced as BRST doublets, i.e.,
\begin{equation}
\s\lambda = J\,,\quad \mathrm{and} \quad \s J = 0\,.
\label{model8}
\end{equation}
Hence, the Euclidean Yang-Mills action quantized in the MAG and in the presence of the operator \eqref{model7} has the following action,
\begin{equation}
S^\alpha_{\rm MAG} = S_{\rm YM} + S_{\rm FP} + \s\int_x \lambda\, \Bigg(\EuScript{O}(\alpha)+\frac{\xi}{2}J\Bigg)\,.
\label{model9}
\end{equation}
The parameter $\xi$ is the so-called LCO-parameter and it is necessary to absorb divergences arising from correlation functions of $\EuScript{O}(\alpha)$. Upon the action of the BRST operator on the last term in eq.~\eqref{model9} one obtains,
\begin{eqnarray}
S^\alpha_{\rm MAG} &=& S_{\rm YM} + S_{\rm FP} + \int_x \Bigg[ J\EuScript{O}(\alpha)+\frac{\xi}{2}J^2\nonumber\\
&+&\lambda\, \Big(A^a_\mu \EuScript{D}^{ab}_\mu c^b+\alpha i b^a c^a - \alpha g \epsilon^{ab}\bar{c}^a c^b c\Big)\Bigg]\,.
\label{model10}
\end{eqnarray}
The addition of the sources $(\lambda,J)$ has the purpose of introducing the ghost-gluon operator in a BRST-invariant fashion into the initial action. Yet the sources should attain their physical values in concrete computations, i.e., $\lambda = 0$ and $J = M^2$ with $M^2$ being a mass-squared parameter. Thus, in the physical limit, the operator $\EuScript{O}(\alpha)$ provides an effective mass term for the off-diagonal gluons and ghosts. Remarkably, the term
\begin{equation}
S_{M^2} = M^2 \int_x\Bigg(\frac{1}{2}A^a_\mu A^a_\mu- \alpha\, \bar{c}^a c^a\Bigg)\,,
\label{model11}
\end{equation}
is BRST-invariant on-shell.

The condensation of the operator \eqref{model7} engenders a dynamical mass generation for the off-diagonal gluons and ghosts. Such a feature is very welcome for the Abelian dominance paradigm since such a mass will decouple the off-diagonal degrees of freedom for sufficiently low momentum scales and lead to an effective Abelian theory in the deep infrared. Yet the MAG is not a complete gauge-fixing due to the Gribov problem \cite{Gribov:1977wm,Singer:1978dk}. Even after the complete implementation of the MAG, gauge field configurations which satisfy the MAG condition and are connected by gauge transformations exist. The Gribov problem in the MAG was investigated in, e.g.,  \cite{Capri:2005tj,Capri:2006cz,Capri:2008ak,Capri:2008vk,Capri:2010an,Gongyo:2013rua}. In the next subsection, we provide a quick overview of such an issue and explain how the removal of Gribov copies affects the dynamics of the Abelian components of the gauge field in the infrared. 

%-------------------------------------------------------
\subsection{Infinitesimal Gribov copies in the MAG \label{Sec:Model.2}}
%-------------------------------------------------------

Gribov copies are defined by normalizable field configurations that satisfy the gauge condition and are connected by a gauge transformation. In the MAG this would correspond to the following statement: Given the field configuration $(A^a_\mu,A_\mu)$ that satisfies
\begin{equation}
\EuScript{D}^{ab}_{\mu}(A)A^{b}_\mu = 0\,, \quad {\rm and} \quad \partial_\mu A_\mu = 0\,,
\label{gcMAG.1}
\end{equation}
a configuration $(\tilde{A}^a_\mu, \tilde{A}_\mu)$ is a Gribov copy associated with $(A^a_\mu,A_\mu)$ if
\begin{equation}
\EuScript{D}^{ab}_{\mu}(\tilde{A})\tilde{A}^{b}_\mu = 0\,, \quad {\rm and} \quad \partial_\mu \tilde{A}_\mu = 0\,,
\label{gcMAG.2}
\end{equation}
and $(\tilde{A}^a_\mu, \tilde{A}_\mu)$ is related to $({A}^a_\mu, {A}_\mu)$ by a gauge transformation. In particular, for infinitesimal gauge transformations generated by $(\xi^a,\xi)$,
\begin{eqnarray}
\tilde{A}^a_\mu &=& A^a_\mu - \EuScript{D}^{ab}_\mu (A) \xi^b - g\epsilon^{ab}A^b_\mu \xi\,, \nonumber\\
\tilde{A}_\mu &=& A_\mu - \partial_\mu \xi - g\epsilon^{ab}A^a_\mu \xi^b\,.
\label{gcMAG.3}
\end{eqnarray}
Plugging equations \eqref{gcMAG.3} into \eqref{gcMAG.2} leads to
\begin{eqnarray}
-\EuScript{D}^{ac}_\mu \EuScript{D}^{cb}_\mu\xi^b - g^2\epsilon^{ac}\epsilon^{bd}A^c_\mu A^d_\mu \xi^b &=& 0\,, \nonumber\\
-\partial^2 \xi - g\epsilon^{ab}\partial_\mu (A^a_\mu \xi^b) &=& 0\,.
\label{gcMAG.4}
\end{eqnarray}
Therefore if equations \eqref{gcMAG.4} have non-trivial solutions, then there exists infinitesimal Gribov copies $(\tilde{A}^a_\mu, \tilde{A}_\mu)$ generated by $(\xi^a,\xi)$. In practice, the second equation in \eqref{gcMAG.4} expresses the Abelian component $\xi$ as a function of the non-Abelian components of the gauge field and the gauge parameter of the infinitesimal transformation. Hence, if the first equation is solved and since it does not involve $\xi$, the second one automatically constrains the form of $\xi$. As such, for the analysis of the existence of infinitesimal Gribov copies in the MAG, investigating the existence of normalizable solutions to the first equation in \eqref{gcMAG.4} is what effectively matters. Moreover, the operator that acts on $\xi^b$ in the first equation in \eqref{gcMAG.4} is precisely the FP operator in the MAG. This operator is Hermitian and hence has real spectrum. Thus, very much like in the Landau gauge, one can propose a region which is free of infinitesimal Gribov copies, the so-called Gribov region, in which the FP operator is positive and thereby restrict the path integral to this region. Such a restriction can be achieved by two different methods: the implementation of the so-called no-pole condition, similarly to the original work \cite{Gribov:1977wm} in the Landau gauge, or the Horizon condition, similarly to \cite{Zwanziger:1989mf} in the Landau gauge. This was explored in \cite{Capri:2005tj,Gongyo:2013rua}. Effectively, the Gribov region in the MAG satisfies important properties such as: it is bounded by the so-called Gribov horizon in every direction in the off-diagonal components of the gauge field; it is unbounded in the diagonal direction; every configuration inside the Gribov region and sufficiently close to the Gribov horizon has a copy that lies on the other side of the horizon and also close to it, see \cite{Capri:2008vk}. 

The restriction of the path integral to the Gribov region in the MAG was worked out in \cite{Capri:2005tj} at leading order by means of the no-pole condition. Its all-order generalization was proposed by demanding consistency with localizability and renormalizability in \cite{Capri:2006cz}. Its all-order construction based on the Horizon condition was worked out in \cite{Gongyo:2013rua}. As far as we are concerned, a proof that the no-pole condition at all orders is equivalent to the Horizon condition in the MAG was not provided up to now. This is contrast to the situation in the Landau gauge, see, e.g., \cite{Capri:2012wx}.

The restriction of the gauge-fixed YM action in the MAG to the Gribov region is effectively achieved by the introduction of the following non-local term,
\begin{equation}
S_h = \gamma^4 H(A) \equiv g^2\gamma^4\int_x \epsilon^{ab}A_\mu\Big[\mathcal{M}^{-1}\Big]^{ac}\epsilon^{cb}A_\mu\,,
\label{gcMAG.5}
\end{equation}
with $\mathcal{M}^{-1}$ denoting the inverse FP operator and $H(A)$ is known as the horizon function. The mass-squared parameter $\gamma^2$ is known as the Gribov parameter and it is not free but fixed by a gap equation. For further details, we refer the reader to \cite{Capri:2005tj}. Albeit non-local, the horizon function can be localized by the introduction of a suitable set of auxiliary fields. The resulting local action is renormalizable at all orders in perturbation theory and implements the restriction of the path integral to the Gribov region in the MAG thereby removing infinitesimal Gribov copies.

From eq.\eqref{gcMAG.5}, it is clear that the horizon function introduces a mass-like term for the Abelian component of the gauge field. This is a non-perturbative effect arising from the removal of Gribov copies. Yet this is not a standard mass term since it is also momentum-dependent due to the non-local structure of the inverse FP operator. Moreover, the introduction of the localizing auxiliary fields is such that their own dynamics renders the formation of condensates. This is due to dynamical infrared instabilities of the model leading to the RGZ action. If we call the mass-squared parameter associated to this condensate as $\mu^2$, then the tree-level Abelian gluon propagator in the RGZ theory is written as \cite{Capri:2008ak}
\begin{equation}
\langle A_\mu (p) A_\nu (-p) \rangle = \frac{p^2+\mu^2}{p^2 (p^2+\mu^2)+4g^2\gamma^4}\Bigg(\delta_{\mu\nu}-\frac{p_\mu p_\nu}{p^2}\Bigg)\,.
\label{gcMAG.6}
\end{equation}
The tree-level non-Abelian gluon propagator is not affected by the restriction of the path integral to the Gribov region. Hence, if one includes the operator \eqref{model7}, its propagator reads
\begin{equation}
\langle A^a_\mu (p) A^b_\nu (-p) \rangle = \frac{\delta^{ab}}{p^2+M^2}\Bigg(\delta_{\mu\nu}-\frac{p_\mu p_\nu}{p^2}\Bigg)\,.
\label{gcMAG.7}
\end{equation}
Some comments are in order: i) At vanishing momentum, the tree-level Abelian propagator attains a finite value hinting to a massive behavior; ii) The mass parameters that provide such a behavior for the Abelian gauge field propagator arise from the restriction of the path integral to the Gribov region. Thus the removal of infinitesimal Gribov copies and the dynamics of the localizing fields were essential to display such a non-trivial behavior of the propagator of the Abelian components of the gauge field; iii) The diagonal gauge fields have an exactly transverse propagator while the non-diagonal components have a transverse propagator at tree level but such a property is lost at higher orders; iv) Although the Abelian propagator reaches a finite value at vanishing momentum, its momentum dependence is very different from a Yukawa-like propagator. In particular, it can be expressed as a sum of Yukawa propagators as follows, 
\begin{equation}
\langle A_\mu (p) A_\nu (-p) \rangle = \Bigg(\frac{\mathsf{R}_{+}}{p^2 + m^2_{+}}+\frac{\mathsf{R}_{-}}{p^2 + m^2_{-}}\Bigg)\,\EuScript{P}_{\mu\nu}\,,
\label{gcMAG.8}
\end{equation}
with 
\begin{equation}
	\EuScript{P}_{\mu\nu}(p)=\delta_{\mu\nu}-\frac{p_\mu p_\nu}{p^2}
\end{equation}
 standing for the transverse projector. The poles $m^2_{\pm}$ and the residues $\mathsf{R}_{\pm}$ are given by
\begin{equation}
m^2_{\pm} = \frac{1}{2}(\mu^2 \pm \Omega^2)\,,
\label{gcMAG.9}
\end{equation}  
and
\begin{equation}
\mathsf{R}_{\pm} = \frac{1}{2}\Big(1 \mp \frac{\mu^2}{\Omega^2}\Big)\,,
\label{gcMAG.10}
\end{equation}  
with $\Omega^2 = \sqrt{\mu^4 - 16g^2\gamma^4}$. 

With such an understanding of the effects of the removal of infinitesimal Gribov copies in the MAG, it is clear that the infrared dynamics of the Abelian sector is affected by mass parameters that have a geometrical nature due to the restriction of the path integral to the Gribov region. Since the RGZ action is not particularly simple due to the large set of fields introduced to localize the Horizon function \eqref{gcMAG.5} and the complicated set of Feynman rules which involve mixings and the proliferation of new diagrams with respect to standard perturbative YM theories, one might raise a legitimate question of practical importance: Can the effects produced by the removal of infinitesimal Gribov copies be mimicked by a simple mass term for the Abelian sector? Such a strategy was pursued in a in the Landau gauge, see, e.g., \cite{Tissier:2010ts,Tissier:2011ey,Gracey:2019xom} leading to the so-called CF model. It provides a highly effective framework that captures the behavior of correlation functions in YM theories quantized in the Landau gauge without the introduction of extra fields, propagators and vertices. In particular, the low-momentum behavior of the underlying correlation functions in the Landau gauge is well described within perturbation theory. It is clear that this model, i.e., the gauge-fixed YM action in the Landau gauge supplemented by a mass-like term for the gluon, is an effective description. One possible first-principle explanation for the introduction of mass-like terms comes precisely from the elimination of Gribov copies. One might wonder if this is just a peculiarity of the Landau gauge or if in other gauges, the introduction of a mass-like term effectively describes the restriction of the path integral to the Gribov region. In this work we provide the first evidence that such extension of gauge-fixed YM theories constitutes a good candidate to describe the infrared behavior of correlation functions in the MAG. 

%-------------------------------------------------------
\subsection{Curci-Ferrari-like Extension in the MAG \label{Sec:Model.3}}
%-------------------------------------------------------

We define the CF-like model in the MAG by the following action,
\begin{equation}
S = S_{\rm YM} + S_{\rm FP} + S_{M^2} + S_{m^2}\,,
\label{Eq:MM.1}
\end{equation}
with 
\begin{equation}
S_{m^2} = \frac{m^2}{2}\int_x A_\mu A_\mu \,.
\label{Eq:MM.2}
\end{equation}
and $S_{M^2}$ is given by Eq.~\eqref{model11}.
As previously described, the $\alpha$-dependent terms are introduced for renormalization purposes, but one should take $\alpha \to 0$ after the computations of the desired quantity in order to recover the MAG. The tree-level non-Abelian gauge field propagator, in the presence of the parameter $\alpha$ is given by
\begin{equation}
\langle A^a_\mu (p) A^b_\nu (-p) \rangle = \frac{\delta^{ab}}{p^2+M^2}\EuScript{P}_{\mu\nu}(p)+\frac{\alpha \delta^{ab}}{p^2+\alpha M^2}\EuScript{L}_{\mu\nu}(p)\,,
\label{Eq:MM.2.1}
\end{equation}
where
\begin{equation}
	 \EuScript{L}_{\mu\nu}(p)=\frac{p_\mu p_\nu}{p^2}.
\end{equation}
For non-vanishing $\alpha$, the tree-level off-diagonal gluon propagator acquires a longitudinal component as shown in Eq.~\eqref{Eq:MM.2.1}. However, it should be stressed that, although the tree-level longitudinal component drops for $\alpha = 0$, it is not expected that such a propagator remains transverse upon inclusion of quantum corrections. This is due to the non-linear nature of the MAG. As for the Abelian gauge field propagator, it reads
\begin{equation}
\langle A_\mu (p) A_\nu (-p) \rangle = \frac{1}{p^2 + m^2}\EuScript{P}_{\mu\nu}(p)\,.
\label{Eq:MM.3}
\end{equation}
This is rather different from the expression \eqref{gcMAG.8}. Thus, as in the Landau gauge, it is expected that one needs to compute loop corrections to such tree-level propagator in order to have a chance to describe the lattice data in the infrared. We will work out the one-loop correction to the gauge-field two-point functions in the next sections.

The mass-like term \eqref{Eq:MM.2} is not invariant under BRST transformations \eqref{model3}. In fact,
\begin{equation}
\s S_{m^2} = m^2 \int_x \Big(c\,\partial_\mu A_\mu + g\epsilon^{ab}c^a A^b_\mu A_\mu\Big)\,,
\label{Eq:MM.4}
\end{equation}
and thus it is not invariant even on-shell. As such, the mass parameter $m^2$ is not akin to a gauge parameter and can enter physical correlation functions. Consequently, the  action defining the model is not invariant under BRST transformations, i.e., $\s S \neq  0$ (even on-shell). One important remark on this matter is that the RGZ action in the MAG \cite{Capri:2005tj} also displays a BRST-breaking due to the restriction of the path integral to the Gribov region. Such a breaking as in the present case is explicit but soft. In fact, a proposal for the reformulation of the RGZ action in the MAG that is BRST invariant was put forward in \cite{Capri:2015pfa}. 

As discussed in the SubSect.~\ref{Sec:Model.1}, it is possible to introduce the mass-like term for the Abelian components of the gauge field in a BRST-invariant fashion upon inclusion of external sources that form a BRST doublet. In the end of concrete computations, the sources attain their physical values recovering the initial model \eqref{Eq:MM.1}. Given the sources $\tau$ and $\rho$ such that 
\begin{equation}
\s \,\tau = \rho\, \quad {\rm and} \quad \s \,\rho = 0\,,
\label{Eq:MM.5}
\end{equation}
one can write
\begin{eqnarray}
S_{\tau \rho} &=& \s \int_x \frac{\tau}{2} \Big( A_\mu A_\mu + \zeta\, \rho\Big)
= \int_{x} \frac{\rho}{2} \Big( A_\mu A_\mu + \zeta\, \rho\Big)\nonumber\\
&+& \int_x \tau\, \Big(A_\mu \partial_\mu c + g\epsilon^{ab} A^a_\mu A_\mu c^b\Big)\,,
\label{Eq:MM.6}
\end{eqnarray}
where the parameter $\zeta$ is the LCO parameter and plays the analogue role of the parameter $\xi$ in eq.~\eqref{model9}. In the physical limit, the sources attain the corresponding values: $\tau = 0$ and $\rho = m^2$. This is useful, for instance, to prove the renormalizability of the model.

%-------------------------------------------------------
\section{Gluon propagators at one-loop order \label{Sec:GP1L}}
%-------------------------------------------------------

In this section, we discuss the essential elements for the explicit evaluation of the non-Abelian and diagonal gluon propagators at one-loop order. We integrate out the Lautrup-Nakanishi fields $(b^a,b)$ and take the physical limits of the corresponding sources $(\lambda,J,\tau,\rho)$, i.e., $\lambda = \tau = 0$,  $J=M^2$ and $\rho = m^2$. For the integration of the Abelian field $b$ we introduce a quadratic term on $b$ together with a gauge parameter $\beta$ that must be set to zero in order to recover the condition $\partial_\mu A_\mu = 0$. The resulting action is written as
\begin{eqnarray}
S &=& \frac{1}{4}\int_x \Big(F^a_{\mu\nu}F^a_{\mu\nu} + F_{\mu\nu}F_{\mu\nu}\Big)+\frac{1}{2\alpha}\int_x \Big(\EuScript{D}^{ab}_{\mu}A^b_\mu\Big)^2\nonumber\\
&+&\frac{1}{2\beta}\int_x (\partial_\mu A_\mu)^2 + \int_x \bar{c}^a \Big(\EuScript{D}^{ac}_{\mu}\EuScript{D}^{cb}_{\mu}+g^2 \epsilon^{ac}\epsilon^{bd}A^c_\mu A^d_\mu\Big)c^d\nonumber\\
&+& \int_x \bar{c}\,\partial_\mu \Big(\partial_\mu c + g\epsilon^{ab} A^a_\mu c^b\Big)-\frac{\alpha}{2}\int_x g^2 (\bar{c}^a c^a)^2 \nonumber\\
&+& M^2\int_x  \Bigg(\frac{1}{2}A^a_\mu A^a_\mu - \alpha\, \bar{c}^a c^a\Bigg)+\frac{m^2}{2}\int_x A_\mu A_\mu\,.
\label{Eq:GP1L.1}
\end{eqnarray}
In the next subsections, we shall present the one-loop corrections to the off-diagonal and diagonal gluon propagators. Our conventions and Feynman rules are collected in the appendices \ref{C} and \ref{FR}.
\\
%-------------------------------------------------------

\subsection{Renormalization Conditions}

\begin{figure}[t!]
	\centering
	\includegraphics[width=\linewidth]{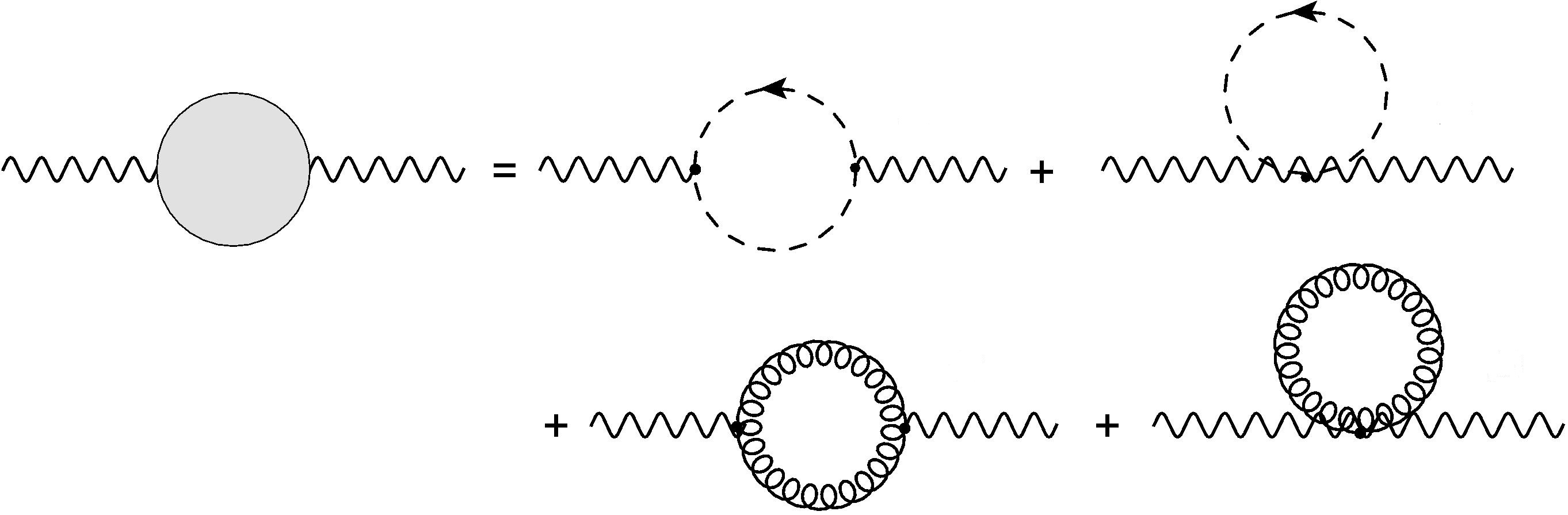}
	\caption{One-loop diagrams that contribute to $\langle A_\mu(p)A_\nu(-p) \rangle$}		
	\label{fig:diagonal_diagrams}
\end{figure}

\begin{figure}[t!]
	\centering
	\includegraphics[width=\linewidth]{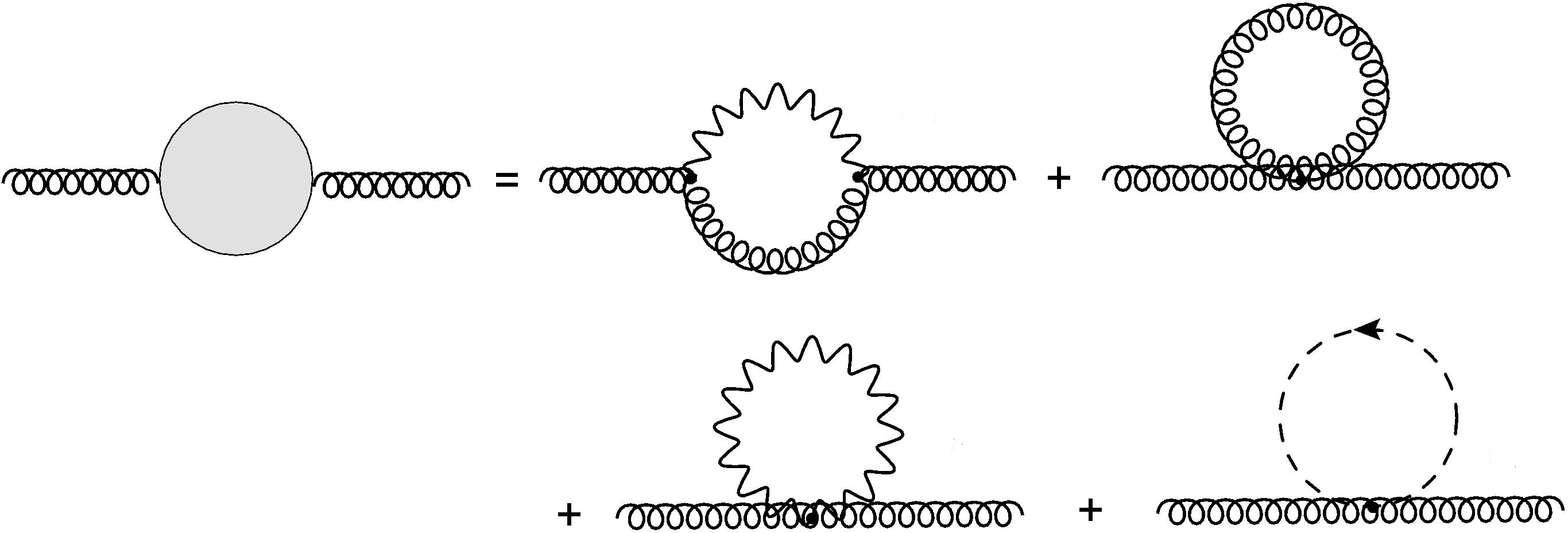}
	\caption{One-loop diagrams that contribute to $\langle A^a_\mu(p)A^b_\nu(-p) \rangle$}
	\label{fig:offdiagonal_diagrams}
\end{figure}

We calculate the diagonal gauge-field propagator
\begin{eqnarray}
	G_{\mu\nu}(p)= \langle A_{\mu}(p) A_{\nu}(-p) \rangle 
\end{eqnarray}
and the off-diagonal propagator 
\begin{eqnarray}
	\mathbb{G}_{\mu\nu}^{ab}(p)= \langle A^a_{\mu}(p) A^b_{\nu}(-p) \rangle 
\end{eqnarray}
up to one-loop order. This requires the calculation of the Feynman diagrams collected in Figs.~\ref{fig:diagonal_diagrams} and \ref{fig:offdiagonal_diagrams}. 

The tree-level propagators and vertex functions necessary for this calculation are set out in Appendix \ref{FR}. The most general structure of both diagonal and off-diagonal propagators is, schematically, 
\begin{eqnarray}
	\mathcal{G}_{\mu\nu}(p)= \left(\delta_{\mu\nu}-\frac{p_{\mu}p_{\nu}}{p^2}\right) \mathcal{G}_{\rm T} (p)+ \frac{p_{\mu}p_{\nu}}{p^2} \mathcal{G}_{\rm L} (p)
	\label{genprop}
\end{eqnarray}
with $\mathcal{G}_{\rm T,L}$ being scalar functions representing the transverse and longitudinal parts of the propagators, respectively. Calligraphic letters are employed to represent that \eqref{genprop} is valid both for the non-Abelian and diagonal propagators with the necessary adjustments. The tree-level expressions for $\mathbb{G}^{ab}_{\mu\nu}$ and $G_{\mu\nu}$ are given in Eq.~\eqref{Eq:MM.2.1} and \eqref{Eq:MM.3}, respectively.  It follows from the Landau-type condition for the diagonal component of the gauge field that the longitudinal part of the diagonal gluon propagator exactly vanishes, that  is 
\begin{equation}
	G_{\rm L} (p)=0.	
\end{equation}

Thus we have three non-trivial form factors to be computed: $G_{\rm T}$ and $\mathbb{G}_{\rm T,L}$. Notice that the tree-level contribution of $\mathbb{G}_{\rm L}$ is zero for $\alpha =0$, see Eq.~\eqref{Eq:MM.2.1}. For renormalization purposes, we keep $\alpha$ and just take the limit $\alpha \to 0$ at the very end of the computation. After resummation, the form factors look like
\begin{align}
	G_{\rm T} (p) &= \frac{1}{p^2+m^2-\Pi^{d}_{\rm T}(p^2)}\,, \label{7a}\\
	\mathbb{G}_{\rm T} (p) &= \frac{1}{p^2+M^2-\Pi^{\rm off}_{\rm T} (p^2)}\,, \label{7b}\\
	\mathbb{G}_{\rm L} (p) &= \frac{\alpha}{p^2+\alpha M^2 - \Pi^{\rm off}_{\rm L} (p^2)}\,,
	\label{7c}
\end{align}
with $\Pi (p^2)$ representing the corresponding self-energies. Employing dimensional regularization with $d=4-\epsilon$, we get the following divergent parts of the one-loop corrections:

%\begin{widetext}
	\begin{align}
		\Pi_{\rm T}^{\rm d}(p^2) \Big|^{\rm 1-loop}_{\text{div}}&= \frac{g^2}{(4\pi)^2 \epsilon}\frac{44p^2}{3}\nonumber\\
		\Pi_{\rm T}^{\rm off}(p^2)\Big|^{\rm 1-loop}_{\text{div}}&= \frac{g^2}{(4\pi)^2 \epsilon}\Bigg[\Big(\frac{17}{3}-\alpha\Big)p^2-\frac{3}{2}m^2(3+\alpha)\nonumber\\
		&-3M^2(1+\alpha)\Bigg]\nonumber\\
		\Pi_{l}^{\rm off}(p^2)\Big|^{\rm 1-loop}_{\text{div.}}&= \frac{g^2}{(4\pi)^2 \epsilon}\Bigg[\Big(\alpha +3+\frac{6}{\alpha}\Big) p^2-\frac{3}{2}  m^2(3\alpha+\alpha^2 )\nonumber\\
		&-3 M^2(\alpha +\alpha^2) \Bigg]
		\label{7}
	\end{align}
%\end{widetext}
Notice that, all divergences are local in the external momentum, $p$, and in the mass parameters, $m^2$ and $M^2$.

All of these divergences can be absorbed by $Z$-factors relating the bare fields and parameters to their renormalized counterparts. Here, we implement the same Momentum Subtraction (MOM) like renormalization conditions as those used in \cite{Tissier:2010ts, Tissier:2011ey} for the one-loop calculation in the Landau gauge-fixed CF model for better comparison with lattice data. They are
\begin{align}
	\Pi(p^2=\mu^2)&=0\,\quad {\rm and} \quad \Pi(p^2=0)=0,
\end{align}
for all $\Pi(p^2)$ given in Eq.~\eqref{7a}, \eqref{7b} and \eqref{7c}. The renormalization factors are defined by
\begin{align}
	A_{\mu}&=Z^{1/2}_A A_{R\mu}, \,\,\,     A^a_{\mu}=Z^{1/2}_{A^a} A^a_{R\mu}, \,\,\, \alpha= Z_{\alpha}\alpha_R \nonumber\\
	m^2&= Z_{m^2} m_R^2, \,\,\,  M^2= Z_{M^2} M_R^2,
\end{align}
and from Eq.~\eqref{7} we can read off their divergent part
\begin{align}
	Z_A&= 1+\frac{g^2}{(4\pi)^2 \epsilon}\frac{44}{3}\,,\nonumber\\
	Z_{A^a}&=1+\frac{g^2}{(4\pi)^2 \epsilon}\left(\frac{17}{3}-\alpha\right)\,,\nonumber\\
	Z_{\alpha}&=1+ \frac{g^2}{(4\pi)^2 \epsilon} \left(\frac{8}{3}-2\alpha -\frac{6}{\alpha}\right)\,,\nonumber\\
	Z_{M^2}&=1-\frac{g^2}{(4\pi)^2 \epsilon}\left(\frac{3}{2}(\alpha +3) \frac{m^2}{M^2}+\frac{26}{3}+2\alpha\right)\,,\nonumber\\
		Z_{m^2}&= 1\,.
\end{align}
Notice that the parameter $m^2$ does not get renormalized up to first order in $g^2$. However, it does appear in the renormalization of $M^2$. The renormalization factors $Z_A$, $Z_{A^a}$ and $Z_{\alpha}$ coincide with those of \cite{Morozov:2005kh}, where the authors considered the MAG without any masses, i.e. $m=M=0$. This means that the addition of a mass term does not change the renormalization of the original fields and parameters, something that was also found in the Landau gauge \cite{Tissier:2010ts,Tissier:2011ey}.

%%%%%%%%%%%%%%%%%%%%%
\subsection{Results}  
%%%%%%%%%%%%%%%%%%%%%

After renormalizing the fields and parameters, we can take $\alpha \rightarrow 0$ which corresponds to the ``true" MAG. The renormalized functions $G_{\rm T} (p^2)$ and $\mathbb{G}_{\rm T} (p^2)$ then read

\begin{widetext}
	\begin{equation}
	\begin{aligned}
	\frac{G_{\rm T}^{-1}(p)}{m^2}
	&= t+1
	- \frac{g^2 t}{576\pi^2}\Bigg\{  - 288 s^{-1}+
	3\left(s^2-4s-4\right)\ln(s) -6\left(s^2-9s+\frac{1}{s}-9\right)\ln(1+s) \\
	&\qquad -\frac{3\sqrt{s(s+4)}\left(s^3-16s^2-68s+48\right)}
	{s^2}\ln\!\Big(\frac{s-\sqrt{s(s+4)}+2}{2}\Big)- (p^2 \rightarrow \mu^2)\Bigg\} \nonumber\\
		\frac{(\mathbb{G}_{\rm T})^{-1}(p)}{M^2} &=s+1-\frac{g^2 s}{192 \pi^2} \Bigg\{\frac{s^2 (-327 t-36)-6 s t (47 t-19)-48 t^2}{6 s^2 t^2}-\frac{\left((s+1)^3 \left(s^2-10 s+1\right) t\right)}{s^4}\ln (s+1)\\
		&+ \left(\frac{((s-10) s+1) (s+1)^3 t}{2s^4}+\frac{7}{2s^3}+\frac{13 s^2-25 s+13}{s t^2}+\frac{25}{s^2}-\frac{13}{s^2 t}-\frac{s}{2t^4}+\frac{7 (s-1)}{2t^3}+\frac{36}{t-s}+\frac{7 s}{2}\right)\ln{\frac{s}{t}}\\
		&-\frac{\left(\left(t^3-9 t^2-9 t+1\right) (s t+s+t)^2\right)}{s t^4} \ln(t+1)+\frac{\left((s+1)^2 t\right) }{s}\ln (t)	\\
		&+\frac{\left(\left(s^2 (t+1)^2+2 s (t-1) t+t^2\right)^{3/2} \left(s^2 ((t-10) t+1)-10 s (t-1) t+t^2\right)\right) }{2 s^4 t^4}\\
		&\times \ln \left(\frac{\sqrt{s^2 (t+1)^2+2 s (t-1) t+t^2}+s t+s+t}{-\sqrt{s^2 (t+1)^2+2 s (t-1) t+t^2}+s t+s+t}\right)- (p^2 \rightarrow \mu^2)	\Bigg\}
	\end{aligned}
	\end{equation}
\end{widetext}
with $s=p^2/M^2$ and $t=p^2/m^2$. We checked that for $m=M=0$ the results, before implementing our renormalization conditions, coincide with the results of \cite{Morozov:2005kh}. As it was reported in \cite{Morozov:2005kh}, in the massless case the longitudinal correction  $\Pi_{\rm L}^{\rm off}$ blows up, while the tree-level propagator vanishes. We found that in the case of finite $m,M$, this effect remains, making an analysis up to first loop order of the longitudinal propagator impossible in the $\alpha \rightarrow 0$ limit. In contrast, lattice simulations of \cite{Bornyakov:2003ee} show a finite longitudinal off-diagonal propagator, saturating in the deep infrared, see Figure \ref{prop}. 

%%%%%%%%%%%%%%%%%%%%%%%%%%
\subsection{Comparison with the lattice results }  
%%%%%%%%%%%%%%%%%%%%%%%%%%

\begin{figure}[t!]
	\includegraphics[width=\linewidth]{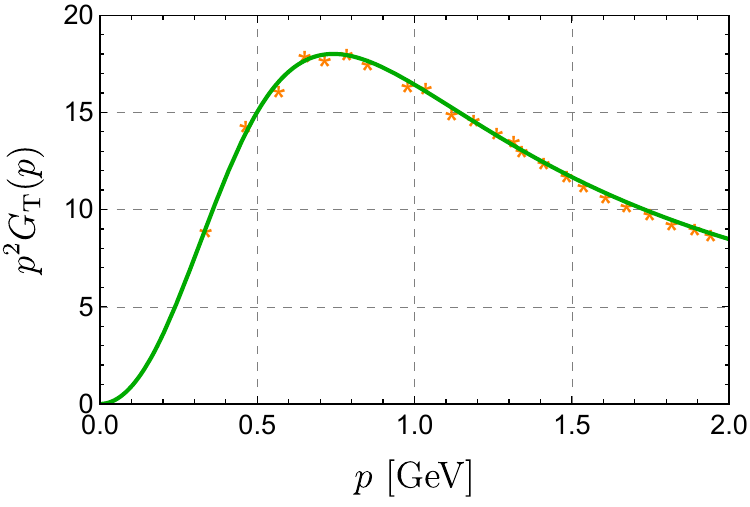}
	\caption{Gluon dressing of the diagonal transverse propagator in the MAG for the $SU(2)$ gauge group. The results of the present work (Green line) are compared with lattice data of \cite{Bornyakov:2003ee} (Orange stars). }
	\label{dressdiagonal}
\end{figure}

\begin{figure}[t!]
	\includegraphics[width=\linewidth]{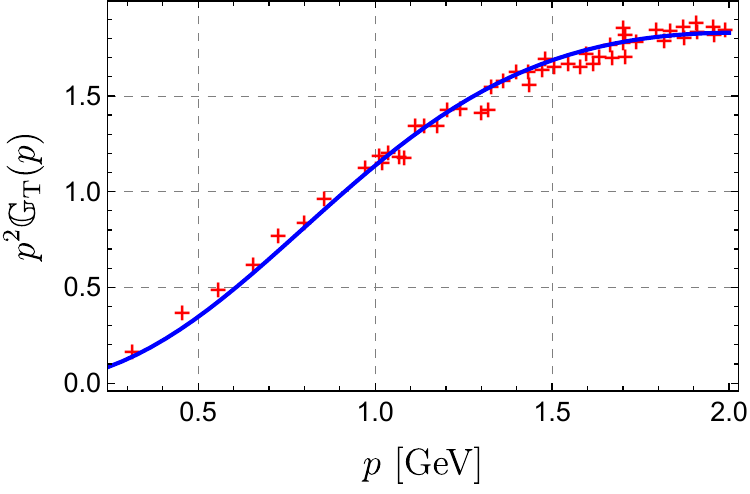}
	\caption{Gluon dressing of the off-diagonal transverse propagator in the MAG for the $SU(2)$ gauge group. The results of the present work (Blue line) are compared with lattice data of \cite{Bornyakov:2003ee} (Red crosses).}
	\label{dressoffdiagonal}
\end{figure}

\begin{figure}[t!]
	\includegraphics[width=\linewidth]{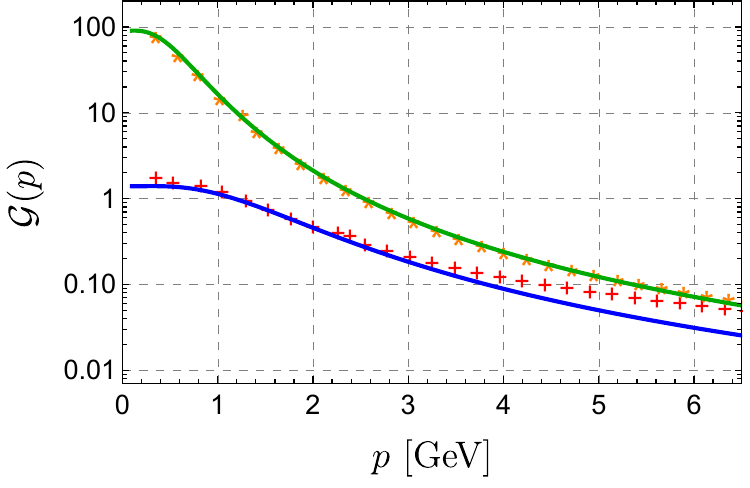}
	\caption{Transverse gluon propagator in the MAG for the $SU(2)$ gauge group. The results of the present work are compared with lattice data of \cite{Bornyakov:2003ee}. For the diagonal transverse propagator: Green line (present work) and Orange stars (lattice data). For the off-diagonal transverse propagator: Blue line (present work) and Red crosses (lattice data). }
	\label{prop}
\end{figure}

We compare our results with those obtained by lattice simulation in \cite{Bornyakov:2003ee}. In Figure \ref{dressdiagonal} and Figure \ref{dressoffdiagonal}, we fit the dressing functions $p^2 G_{\rm T} (p)$ and $p^2 \mathbb{G}_{\rm T} (p)$ with the lattice data from the simulations reported in \cite{Bornyakov:2003ee}. The fits are obtained with the constraint $M>m$. We find that the best parameter fits are $Z_{\rm diag}=13.4$, $Z_{\rm off}=3.9$, $g=8.1$, $M=1.7$ GeV, $\mu=1.4$ GeV and $m=0.4$ GeV. The parameters $Z_{\rm diag}$ and $Z_{\rm off}$ are global multiplicative parameters necessary to adjust to the normalization of lattice data. Notice that these values are within the same order as those obtained in the Landau gauge \cite{Tissier:2010ts, Tissier:2011ey}. We find a satisfactory fit in the IR regime, as can be seen from Figure \ref{dressdiagonal} and Figure \ref{dressoffdiagonal}. We also compared our fitting with the lattice data on the transverse components of the propagators, which was given in \cite{Bornyakov:2003ee} for larger values of $p$, see Figure \ref{prop}. It is clear that for larger values of $p$, the fitting does not work that well. In this case, we certainly need a renormalization group improvement before drawing a conclusion.

\begin{figure}[t!]
	\includegraphics[width=\linewidth]{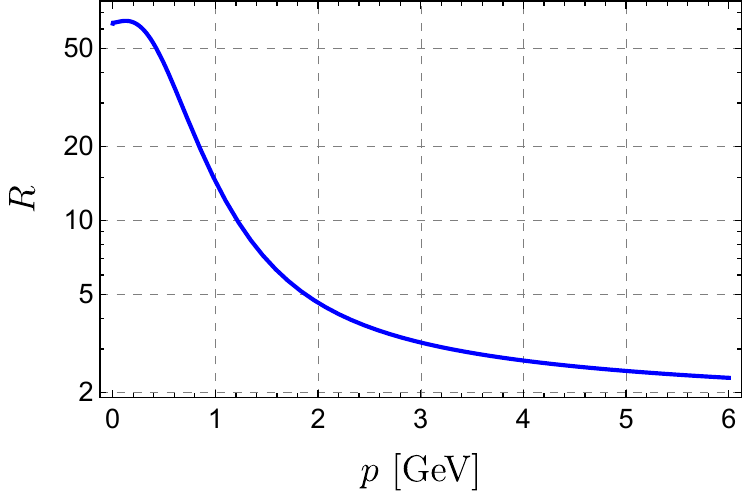}
	\caption{Ratio of the transverse diagonal and off-diagonal propagators.}
	\label{ratio}
\end{figure}

From these results, it is clear that neither propagator vanishes or diverges in the deep IR, i.e., $p \rightarrow 0$. However, at low momenta the diagonal transverse propagator is much larger than the off-diagonal propagator. This can be interpreted as a hint toward Abelian dominance. In Figure \ref{ratio}, we compare the ratio of $G_{\rm T}$ and $\mathbb{G}_{\rm T}$. As can be seen, the suppression of the off-diagonal propagator with respect to the diagonal propagatot increases as the momentum decreases. Such an observation can be taken as an indication that, in the IR, the diagonal gluons are the dominant degrees of freedom. Notice that since for our fitting parameters $M>m$, this effect is already present at tree-level. 

%-------------------------------------------------------
\section{Conclusions \label{Sec:Conclusions}}
%-------------------------------------------------------

In this work, we have shown that the CF-like model quantized in the MAG reproduces the transverse components of the gluonic two-point correlation functions in $SU(2)$ YM theories for small momenta. We calculated up to first-loop order the off-diagonal gluon propagator and the diagonal gluon propagator, $\langle A^a_\mu(p) A^b_\nu (-p) \rangle$ and $\langle A^3_\mu(p) A^3_\nu (-p) \rangle$, respectively. We employed Dimensional Regularization and MOM-type renormalization conditions for the three non-zero form factors. By fitting the available free parameters, we are able to obtain a rather satisfactory description of the transverse components of the propagators obtained in lattice simulations in the IR regime. This seems promising, since there is room for many improvements as the inclusion of the runnings of the coupling and anomalous dimensions. Unfortunately, the longitudinal part of the off-diagonal component is inaccessible in our approach, simply because in the limit $\alpha \rightarrow 0$ the tree-level contribution vanishes and, at the limit $p \rightarrow 0$, the one-loop correction blows up. 

The results obtained in this work are comparable to those in the CF model in the Landau gauge, which has been highly successful in reproducing a range of non-perturbative features of pure Yang-Mills theories and QCD. Of course, many improvements were achieved in the Landau gauge and such an effective model has shown to be rather robust. It is unknown if the same will hold in the MAG and further investigations are necessary. The importance of this result is two-fold. First, it reinforces the status of the CF model as a relevant phenomenological framework for low-energy QCD, demonstrating that its effectiveness is potentially not limited to the Landau gauge. Second, although the MAG is undoubtedly more involved than the Landau gauge from a computational standpoint, it brings to light certain features of Yang–Mills theories, e.g. Abelian dominance, that remain obscure in the Landau gauge. 

These results are encouraging for future investigations of other non-perturbative features of gauge-fixed correlation functions in the MAG. For example, one can consider a Renormalization Group (RG) improvement following the IR-safe RG scheme proposed in \cite{Tissier:2010ts, Tissier:2011ey}. We expect such an improvement leads to better results for larger values of $p$, which, for our simple approach, are not very satisfying. It remains to be investigated how to consistently account for the longitudinal sector of the off-diagonal gauge-field propagator which seems difficult to be accessed meaningfully at one-loop order in the IR. In addition, lattice data is available for $SU(3)$ YM theories, see \cite{Gongyo:2013sha}. Studying the gluon propagator up to first loop order in the $SU(3)$ CF-like model with MAG is a non-trivial extension of the $SU(2)$ calculations presented here, due to the fact that there are now two Abelian components to the gauge field, $A^3_\mu$ and $A^8_\mu$, which means that the residual symmetry is an $U(1) \times U(1)$ symmetry and more vertices are involved. One topic that is particularly scarce in the MAG literature is the inclusion of matter and, in the present framework, we can investigate the coupling of scalars and fermions which are of great phenomenological importance.

%-------------------------------------------------------
\section*{Acknowledgments}
%-------------------------------------------------------

We would like to thank J.~P.~Vaz e Silva for enlightening discussions. ADP acknowledges financial support from CNPq under the grant PQ-2 (312211/2022-8) and FAPERJ under the “Jovem Cientista do Nosso Estado” program (E-26/205.924/2022 and E-26/204.457/2025). GP is a FAPERJ postdoctoral fellow in the PÓS DOUTORADO NOTA 10 program under the contracts E-26/205.924/2022 and E-26/205.925/2022. DvE is a FAPESP postdoctoral fellow under the contract 2023/03722-9. SPS acknowledges CNPq for financial support under the contract 302991/2024-7. 

%%%%%%%%%%%%%%%%%%%%%%
%%%%%%APPENDICES%%%%%%%%%%
%%%%%%%%%%%%%%%%%%%%%%

\appendix

\begin{widetext}
%-------------------------------------------------------
\section{Conventions \label{C}}
%-------------------------------------------------------

In this paper we work on four-dimensional flat Euclidean space and adopt the following convention for Fourier transform for any field $\Phi (x)$,
\begin{equation}
\Phi (x) = \int \frac{\mathrm{d}^4p}{(2\pi)^4}{\rm e}^{-ix\cdot p}\tilde{\Phi}(p)\,.
\label{Ap:Conv.1}
\end{equation}
For convenience, we shall employ the short-hand notation $\int \frac{\mathrm{d}^4p}{(2\pi)^4} = \int_p$ or $\int \frac{\mathrm{d}^dp}{(2\pi)^d} = \int_{p^d}$ when an analytic extension of the space dimensionality is employed. 

%-------------------------------------------------------
\section{Feynman Rules \label{FR}}
%-------------------------------------------------------

\subsection{Propagators}

The propagators of the theory are determined by the quadratic part of $S$, namely,
\begin{eqnarray}
	S_{\text{quad}} &=& \int_x \left[\frac{1}{4}(\partial_{\mu} A^a_{\nu} - \partial_{\nu} A^a_{\mu})^2+\frac{1}{4}(\partial_{\mu} A_{\nu} - \partial_{\nu} A_{\mu})^2+\frac{1}{2\alpha}(\partial_{\mu}A_{\mu}^a)^2+\frac{1}{2\beta}(\partial_{\mu}A_{\mu})^2+\frac{M^2}{2}A_{\mu}^aA_{\mu}^a+\frac{m^2}{2}A_{\mu}A_{\mu}\right] \nonumber \\ 
	& & + \int_x\left( \bar{c}^a\partial^2 c^a-\alpha M^2 \bar{c}^ac^a+\bar{c}\partial^2 c \right).
\end{eqnarray}
Notice that the bosonic sector can be written as $\int_{x,y} \frac{1}{2}\Phi^T (x) \Delta (x-y) \Phi (y)$, where 
\begin{equation}
	\Phi = \begin{pmatrix}
		A^a_{\mu}\\
		A_{\mu} \\
	\end{pmatrix}
\end{equation}
and
\begin{equation}
	\Delta (x-y) = \begin{pmatrix}
		\delta^{ab}\left[-\delta_{\mu\nu}(\partial^2-M^2)+(1-\frac{1}{\alpha})\partial_{\mu}\partial_{\nu}\right] & 0 \\
		0 & \left[-\delta_{\mu\nu}(\partial^2-m^2)+(1-\frac{1}{\beta})\partial_{\mu}\partial_{\nu}\right]
	\end{pmatrix}
	.
\end{equation}
We obtain the propagators by inverting $\Delta(x-y)$, which is not a difficult task, as there are no mixing terms. By doing that and in the end taking the limits $\alpha \rightarrow 0$, $\beta \rightarrow 0$, we get
\begin{eqnarray}
	\langle A^a_{\mu}(p)A^b_{\nu}(-p) \rangle &=& \frac{\delta^{ab}}{p^2+M^2}\EuScript{P}_{\mu\nu}(p), \nonumber \\
	\langle A_{\mu}(p)A_{\nu}(-p) \rangle &=& \frac{1}{p^2+m^2}\EuScript{P}_{\mu\nu}(p).
\end{eqnarray} 
Ghost propagators are obtained in a similar way, but taking into account the Grassmannian nature of the fields. It follows that
\begin{eqnarray}
	\langle \bar{c}^a(p) c^b (-p) \rangle & = & \frac{\delta^{ab}}{p^2},\nonumber \\
	\langle \bar{c}(p) c(-p) \rangle & = & \frac{1}{p^2}.
\end{eqnarray}
The diagrammatic notation for the propagators is shown in Figure \ref{fig:propagators}.

\begin{figure}[H]
	\centering
	\includegraphics[width=10cm]{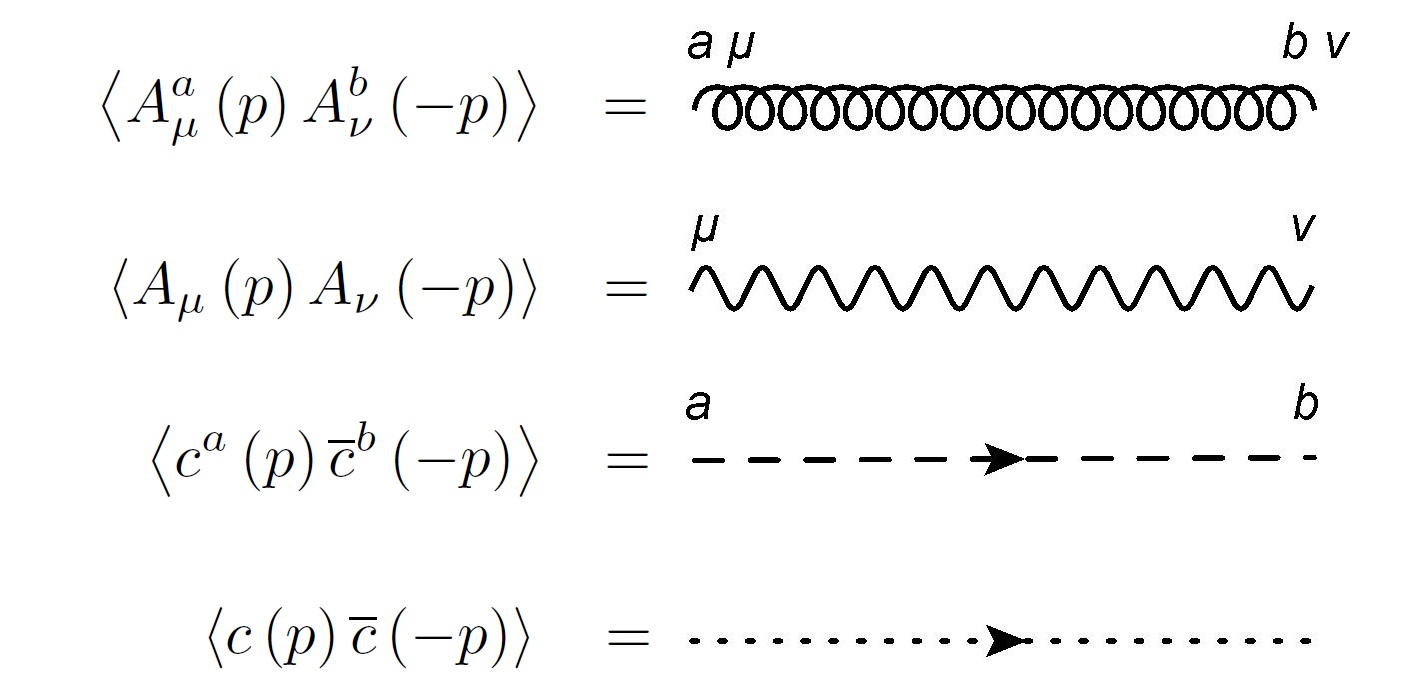}
	\label{fig:propagators}
	\caption{Diagrammatic representation for the propagators.}
\end{figure}

\subsection{Vertices}

The vertices of the theory are determined by the interacting action, which is
\begin{eqnarray}
	S_{\text{int}} &=& S-S_{\text{quad}} \nonumber \\
	               &=& \int_x \left[ g\epsilon^{ab}(\partial_{\mu}A_{\nu}^a)A_{\mu}^b A_{\nu} - g\epsilon^{ab}(\partial_{\mu}A_{\nu}^a)A_{\mu}A_{\nu}^b+g\epsilon^{ab}(\partial_{\mu}A_{\nu})A_{\mu}^b A_{\nu}^c \right. \nonumber \\
	               & & +\frac{g^2}{4}(A_{\mu}^a A_{\mu}^a A_{\nu}^b A_{\nu}^b+2A_{\mu}^aA_{\mu}^aA_{\nu}A_{\nu}-A_{\mu}^aA_{\nu}^aA_{\mu}^bA_{\nu}^b-2A_{\mu}^aA_{\nu}^aA_{\mu}A_{\nu}) \nonumber \\
	               & &-g\epsilon^{ab}\bar{c}^a\partial_{\mu}(A_{\mu}c^b)-g\epsilon^{ab}\bar{c}^aA_{\mu}\partial_{\mu}c^b-g^2\bar{c}^ac^bA_{\mu}A_{\mu}+g^2\epsilon^{ac}\epsilon^{bd}\bar{c}^ac^bA_{\mu}^cA_{\mu}^d \nonumber \\
	               & & \left. +g\epsilon^{ab}\bar{c}\partial_{\mu}(A_{\mu}^ac^b)-\frac{\alpha g^2}{2}\bar{c}^ac^a\bar{c}^bc^b \right].
\end{eqnarray}
A vertex factor, $V_{\phi_{i_1}\dots \phi_{i_n}}(p_1,\dots, p_n)$, corresponding to an interaction between the fields $\phi_{i_1}$,\dots, $\phi_{i_n}$, is defined  in momentum space by 
\begin{eqnarray}
	(2\pi)^d\delta^d(p_1+\dots+p_n)V_{\phi_{i_1}\dots \phi_{i_n}}(p_1,\dots, p_n)&=&-\int_{x_1,\dots,x_n}e^{i(x_1 \cdot p_1+\dots + x_n \cdot p_n)}\left.\frac{\delta^n S_{\text{int}}}{\delta \phi_{i_1}(x_1)\dots \delta \phi_{i_n}(x_n)}\right|_{\phi=0}.
\end{eqnarray}
All vertex factors present in this theory are shown in Figure \ref{fig:vertex_factors}. 

\begin{figure}[H]
	\centering
	\includegraphics[width=15cm]{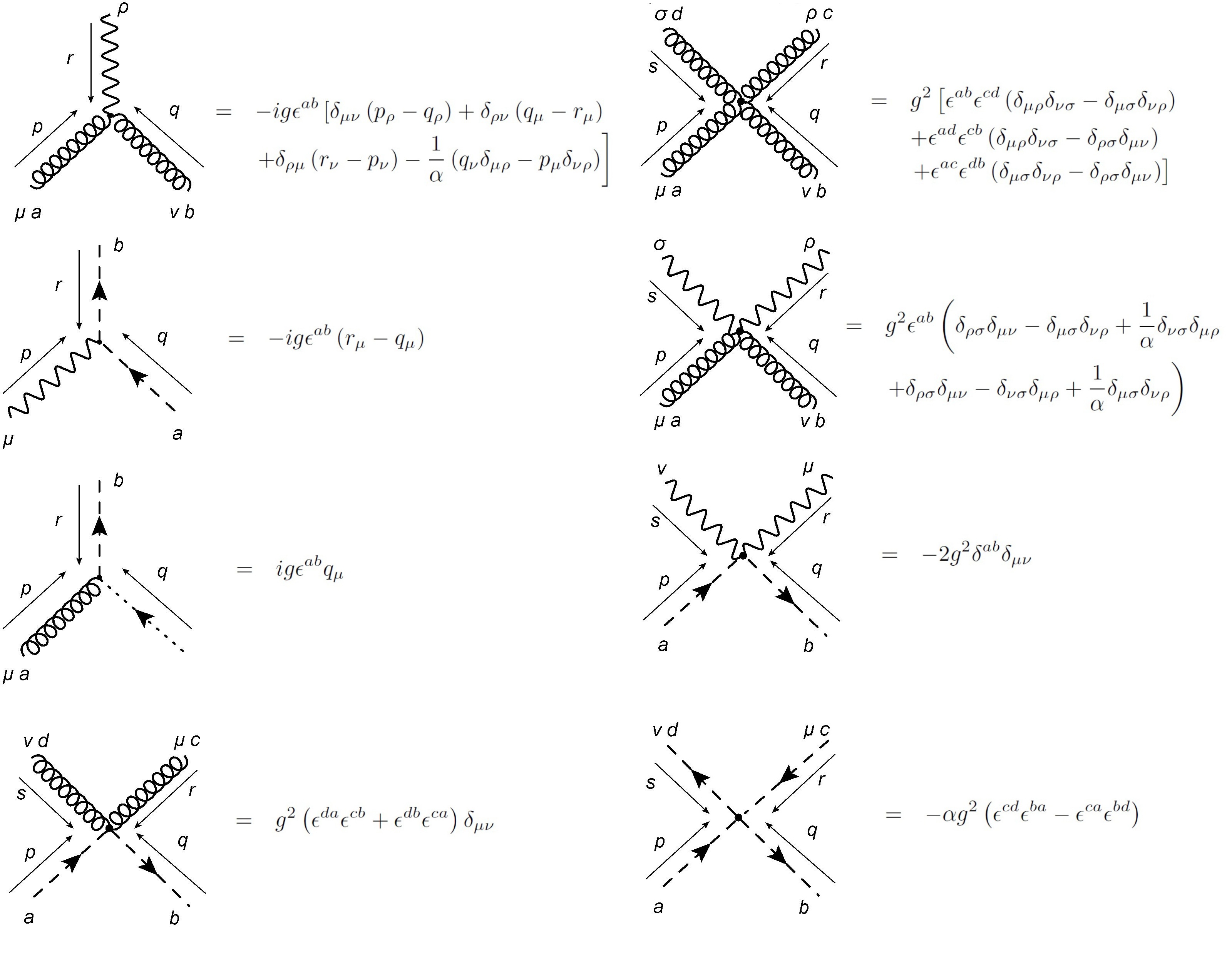}
	\label{fig:vertex_factors}
	\caption{Vertex factors.}

\end{figure}

\end{widetext}

%\newpage

\bibliography{refs}

%apsrev4-2.bst 2019-01-14 (MD) hand-edited version of apsrev4-1.bst
%Control: key (0)
%Control: author (8) initials jnrlst
%Control: editor formatted (1) identically to author
%Control: production of article title (0) allowed
%Control: page (0) single
%Control: year (1) truncated
%Control: production of eprint (0) enabled
\begin{thebibliography}{93}%
\makeatletter
\providecommand \@ifxundefined [1]{%
 \@ifx{#1\undefined}
}%
\providecommand \@ifnum [1]{%
 \ifnum #1\expandafter \@firstoftwo
 \else \expandafter \@secondoftwo
 \fi
}%
\providecommand \@ifx [1]{%
 \ifx #1\expandafter \@firstoftwo
 \else \expandafter \@secondoftwo
 \fi
}%
\providecommand \natexlab [1]{#1}%
\providecommand \enquote  [1]{``#1''}%
\providecommand \bibnamefont  [1]{#1}%
\providecommand \bibfnamefont [1]{#1}%
\providecommand \citenamefont [1]{#1}%
\providecommand \href@noop [0]{\@secondoftwo}%
\providecommand \href [0]{\begingroup \@sanitize@url \@href}%
\providecommand \@href[1]{\@@startlink{#1}\@@href}%
\providecommand \@@href[1]{\endgroup#1\@@endlink}%
\providecommand \@sanitize@url [0]{\catcode `\\12\catcode `\$12\catcode
  `\&12\catcode `\#12\catcode `\^12\catcode `\_12\catcode `\%12\relax}%
\providecommand \@@startlink[1]{}%
\providecommand \@@endlink[0]{}%
\providecommand \url  [0]{\begingroup\@sanitize@url \@url }%
\providecommand \@url [1]{\endgroup\@href {#1}{\urlprefix }}%
\providecommand \urlprefix  [0]{URL }%
\providecommand \Eprint [0]{\href }%
\providecommand \doibase [0]{https://doi.org/}%
\providecommand \selectlanguage [0]{\@gobble}%
\providecommand \bibinfo  [0]{\@secondoftwo}%
\providecommand \bibfield  [0]{\@secondoftwo}%
\providecommand \translation [1]{[#1]}%
\providecommand \BibitemOpen [0]{}%
\providecommand \bibitemStop [0]{}%
\providecommand \bibitemNoStop [0]{.\EOS\space}%
\providecommand \EOS [0]{\spacefactor3000\relax}%
\providecommand \BibitemShut  [1]{\csname bibitem#1\endcsname}%
\let\auto@bib@innerbib\@empty
%</preamble>
\bibitem [{\citenamefont {Faddeev}\ and\ \citenamefont
  {Popov}(1967)}]{Faddeev:1967fc}%
  \BibitemOpen
  \bibfield  {author} {\bibinfo {author} {\bibfnamefont {L.}~\bibnamefont
  {Faddeev}}\ and\ \bibinfo {author} {\bibfnamefont {V.}~\bibnamefont
  {Popov}},\ }\bibfield  {title} {\bibinfo {title} {{Feynman Diagrams for the
  Yang-Mills Field}},\ }\href {https://doi.org/10.1016/0370-2693(67)90067-6}
  {\bibfield  {journal} {\bibinfo  {journal} {Phys. Lett. B}\ }\textbf
  {\bibinfo {volume} {25}},\ \bibinfo {pages} {29} (\bibinfo {year}
  {1967})}\BibitemShut {NoStop}%
\bibitem [{\citenamefont {Sternbeck}\ \emph {et~al.}(2007)\citenamefont
  {Sternbeck}, \citenamefont {von Smekal}, \citenamefont {Leinweber},\ and\
  \citenamefont {Williams}}]{Sternbeck:2007ug}%
  \BibitemOpen
  \bibfield  {author} {\bibinfo {author} {\bibfnamefont {A.}~\bibnamefont
  {Sternbeck}}, \bibinfo {author} {\bibfnamefont {L.}~\bibnamefont {von
  Smekal}}, \bibinfo {author} {\bibfnamefont {D.~B.}\ \bibnamefont
  {Leinweber}},\ and\ \bibinfo {author} {\bibfnamefont {A.~G.}\ \bibnamefont
  {Williams}},\ }\bibfield  {title} {\bibinfo {title} {{Comparing SU(2) to
  SU(3) gluodynamics on large lattices}},\ }\href
  {https://doi.org/10.22323/1.042.0340} {\bibfield  {journal} {\bibinfo
  {journal} {PoS}\ }\textbf {\bibinfo {volume} {LATTICE2007}},\ \bibinfo
  {pages} {340} (\bibinfo {year} {2007})},\ \Eprint
  {https://arxiv.org/abs/0710.1982} {arXiv:0710.1982 [hep-lat]} \BibitemShut
  {NoStop}%
\bibitem [{\citenamefont {Cucchieri}\ and\ \citenamefont
  {Mendes}(2007)}]{Cucchieri:2007md}%
  \BibitemOpen
  \bibfield  {author} {\bibinfo {author} {\bibfnamefont {A.}~\bibnamefont
  {Cucchieri}}\ and\ \bibinfo {author} {\bibfnamefont {T.}~\bibnamefont
  {Mendes}},\ }\bibfield  {title} {\bibinfo {title} {{What's up with IR gluon
  and ghost propagators in Landau gauge? A puzzling answer from huge
  lattices}},\ }\href {https://doi.org/10.22323/1.042.0297} {\bibfield
  {journal} {\bibinfo  {journal} {PoS}\ }\textbf {\bibinfo {volume}
  {LATTICE2007}},\ \bibinfo {pages} {297} (\bibinfo {year} {2007})},\ \Eprint
  {https://arxiv.org/abs/0710.0412} {arXiv:0710.0412 [hep-lat]} \BibitemShut
  {NoStop}%
\bibitem [{\citenamefont {Cucchieri}\ and\ \citenamefont
  {Mendes}(2008)}]{Cucchieri:2007rg}%
  \BibitemOpen
  \bibfield  {author} {\bibinfo {author} {\bibfnamefont {A.}~\bibnamefont
  {Cucchieri}}\ and\ \bibinfo {author} {\bibfnamefont {T.}~\bibnamefont
  {Mendes}},\ }\bibfield  {title} {\bibinfo {title} {{Constraints on the IR
  behavior of the gluon propagator in Yang-Mills theories}},\ }\href
  {https://doi.org/10.1103/PhysRevLett.100.241601} {\bibfield  {journal}
  {\bibinfo  {journal} {Phys. Rev. Lett.}\ }\textbf {\bibinfo {volume} {100}},\
  \bibinfo {pages} {241601} (\bibinfo {year} {2008})},\ \Eprint
  {https://arxiv.org/abs/0712.3517} {arXiv:0712.3517 [hep-lat]} \BibitemShut
  {NoStop}%
%%CITATION = ARXIV:0712.3517;%%
\bibitem [{\citenamefont {Bornyakov}\ \emph {et~al.}(2009)\citenamefont
  {Bornyakov}, \citenamefont {Mitrjushkin},\ and\ \citenamefont
  {Muller-Preussker}}]{Bornyakov:2008yx}%
  \BibitemOpen
  \bibfield  {author} {\bibinfo {author} {\bibfnamefont {V.~G.}\ \bibnamefont
  {Bornyakov}}, \bibinfo {author} {\bibfnamefont {V.~K.}\ \bibnamefont
  {Mitrjushkin}},\ and\ \bibinfo {author} {\bibfnamefont {M.}~\bibnamefont
  {Muller-Preussker}},\ }\bibfield  {title} {\bibinfo {title} {{Infrared
  behavior and Gribov ambiguity in SU(2) lattice gauge theory}},\ }\href
  {https://doi.org/10.1103/PhysRevD.79.074504} {\bibfield  {journal} {\bibinfo
  {journal} {Phys. Rev. D}\ }\textbf {\bibinfo {volume} {79}},\ \bibinfo
  {pages} {074504} (\bibinfo {year} {2009})},\ \Eprint
  {https://arxiv.org/abs/0812.2761} {arXiv:0812.2761 [hep-lat]} \BibitemShut
  {NoStop}%
\bibitem [{\citenamefont {Bogolubsky}\ \emph {et~al.}(2009)\citenamefont
  {Bogolubsky}, \citenamefont {Ilgenfritz}, \citenamefont {Muller-Preussker},\
  and\ \citenamefont {Sternbeck}}]{Bogolubsky:2009dc}%
  \BibitemOpen
  \bibfield  {author} {\bibinfo {author} {\bibfnamefont {I.~L.}\ \bibnamefont
  {Bogolubsky}}, \bibinfo {author} {\bibfnamefont {E.~M.}\ \bibnamefont
  {Ilgenfritz}}, \bibinfo {author} {\bibfnamefont {M.}~\bibnamefont
  {Muller-Preussker}},\ and\ \bibinfo {author} {\bibfnamefont {A.}~\bibnamefont
  {Sternbeck}},\ }\bibfield  {title} {\bibinfo {title} {{Lattice gluodynamics
  computation of Landau gauge Green's functions in the deep infrared}},\ }\href
  {https://doi.org/10.1016/j.physletb.2009.04.076} {\bibfield  {journal}
  {\bibinfo  {journal} {Phys. Lett.}\ }\textbf {\bibinfo {volume} {B676}},\
  \bibinfo {pages} {69} (\bibinfo {year} {2009})},\ \Eprint
  {https://arxiv.org/abs/0901.0736} {arXiv:0901.0736 [hep-lat]} \BibitemShut
  {NoStop}%
%%CITATION = ARXIV:0901.0736;%%
\bibitem [{\citenamefont {Dudal}\ \emph {et~al.}(2010)\citenamefont {Dudal},
  \citenamefont {Oliveira},\ and\ \citenamefont {Vandersickel}}]{Dudal:2010tf}%
  \BibitemOpen
  \bibfield  {author} {\bibinfo {author} {\bibfnamefont {D.}~\bibnamefont
  {Dudal}}, \bibinfo {author} {\bibfnamefont {O.}~\bibnamefont {Oliveira}},\
  and\ \bibinfo {author} {\bibfnamefont {N.}~\bibnamefont {Vandersickel}},\
  }\bibfield  {title} {\bibinfo {title} {{Indirect lattice evidence for the
  Refined Gribov-Zwanziger formalism and the gluon condensate
  $\langle{A^2}\rangle$ in the Landau gauge}},\ }\href
  {https://doi.org/10.1103/PhysRevD.81.074505} {\bibfield  {journal} {\bibinfo
  {journal} {Phys. Rev. D}\ }\textbf {\bibinfo {volume} {81}},\ \bibinfo
  {pages} {074505} (\bibinfo {year} {2010})},\ \Eprint
  {https://arxiv.org/abs/1002.2374} {arXiv:1002.2374 [hep-lat]} \BibitemShut
  {NoStop}%
\bibitem [{\citenamefont {Maas}(2013)}]{Maas:2011se}%
  \BibitemOpen
  \bibfield  {author} {\bibinfo {author} {\bibfnamefont {A.}~\bibnamefont
  {Maas}},\ }\bibfield  {title} {\bibinfo {title} {{Describing gauge bosons at
  zero and finite temperature}},\ }\href
  {https://doi.org/10.1016/j.physrep.2012.11.002} {\bibfield  {journal}
  {\bibinfo  {journal} {Phys. Rept.}\ }\textbf {\bibinfo {volume} {524}},\
  \bibinfo {pages} {203} (\bibinfo {year} {2013})},\ \Eprint
  {https://arxiv.org/abs/1106.3942} {arXiv:1106.3942 [hep-ph]} \BibitemShut
  {NoStop}%
\bibitem [{\citenamefont {Cucchieri}\ \emph {et~al.}(2012)\citenamefont
  {Cucchieri}, \citenamefont {Dudal}, \citenamefont {Mendes},\ and\
  \citenamefont {Vandersickel}}]{Cucchieri:2011ig}%
  \BibitemOpen
  \bibfield  {author} {\bibinfo {author} {\bibfnamefont {A.}~\bibnamefont
  {Cucchieri}}, \bibinfo {author} {\bibfnamefont {D.}~\bibnamefont {Dudal}},
  \bibinfo {author} {\bibfnamefont {T.}~\bibnamefont {Mendes}},\ and\ \bibinfo
  {author} {\bibfnamefont {N.}~\bibnamefont {Vandersickel}},\ }\bibfield
  {title} {\bibinfo {title} {{Modeling the Gluon Propagator in Landau Gauge:
  Lattice Estimates of Pole Masses and Dimension-Two Condensates}},\ }\href
  {https://doi.org/10.1103/PhysRevD.85.094513} {\bibfield  {journal} {\bibinfo
  {journal} {Phys. Rev. D}\ }\textbf {\bibinfo {volume} {85}},\ \bibinfo
  {pages} {094513} (\bibinfo {year} {2012})},\ \Eprint
  {https://arxiv.org/abs/1111.2327} {arXiv:1111.2327 [hep-lat]} \BibitemShut
  {NoStop}%
\bibitem [{\citenamefont {Oliveira}\ and\ \citenamefont
  {Silva}(2012)}]{Oliveira:2012eh}%
  \BibitemOpen
  \bibfield  {author} {\bibinfo {author} {\bibfnamefont {O.}~\bibnamefont
  {Oliveira}}\ and\ \bibinfo {author} {\bibfnamefont {P.~J.}\ \bibnamefont
  {Silva}},\ }\bibfield  {title} {\bibinfo {title} {{The lattice Landau gauge
  gluon propagator: lattice spacing and volume dependence}},\ }\href
  {https://doi.org/10.1103/PhysRevD.86.114513} {\bibfield  {journal} {\bibinfo
  {journal} {Phys. Rev. D}\ }\textbf {\bibinfo {volume} {86}},\ \bibinfo
  {pages} {114513} (\bibinfo {year} {2012})},\ \Eprint
  {https://arxiv.org/abs/1207.3029} {arXiv:1207.3029 [hep-lat]} \BibitemShut
  {NoStop}%
\bibitem [{\citenamefont {Dudal}\ \emph {et~al.}(2018)\citenamefont {Dudal},
  \citenamefont {Oliveira},\ and\ \citenamefont {Silva}}]{Dudal:2018cli}%
  \BibitemOpen
  \bibfield  {author} {\bibinfo {author} {\bibfnamefont {D.}~\bibnamefont
  {Dudal}}, \bibinfo {author} {\bibfnamefont {O.}~\bibnamefont {Oliveira}},\
  and\ \bibinfo {author} {\bibfnamefont {P.~J.}\ \bibnamefont {Silva}},\
  }\bibfield  {title} {\bibinfo {title} {{High precision statistical Landau
  gauge lattice gluon propagator computation vs.~the
  Gribov\textendash{}Zwanziger approach}},\ }\href
  {https://doi.org/10.1016/j.aop.2018.08.019} {\bibfield  {journal} {\bibinfo
  {journal} {Annals Phys.}\ }\textbf {\bibinfo {volume} {397}},\ \bibinfo
  {pages} {351} (\bibinfo {year} {2018})},\ \Eprint
  {https://arxiv.org/abs/1803.02281} {arXiv:1803.02281 [hep-lat]} \BibitemShut
  {NoStop}%
\bibitem [{\citenamefont {Brambilla}\ \emph {et~al.}(2014)\citenamefont
  {Brambilla} \emph {et~al.}}]{Brambilla:2014jmp}%
  \BibitemOpen
  \bibfield  {author} {\bibinfo {author} {\bibfnamefont {N.}~\bibnamefont
  {Brambilla}} \emph {et~al.},\ }\bibfield  {title} {\bibinfo {title} {{QCD and
  Strongly Coupled Gauge Theories: Challenges and Perspectives}},\ }\href
  {https://doi.org/10.1140/epjc/s10052-014-2981-5} {\bibfield  {journal}
  {\bibinfo  {journal} {Eur. Phys. J. C}\ }\textbf {\bibinfo {volume} {74}},\
  \bibinfo {pages} {2981} (\bibinfo {year} {2014})},\ \Eprint
  {https://arxiv.org/abs/1404.3723} {arXiv:1404.3723 [hep-ph]} \BibitemShut
  {NoStop}%
\bibitem [{\citenamefont {Fischer}\ \emph {et~al.}(2009)\citenamefont
  {Fischer}, \citenamefont {Maas},\ and\ \citenamefont
  {Pawlowski}}]{Fischer:2008uz}%
  \BibitemOpen
  \bibfield  {author} {\bibinfo {author} {\bibfnamefont {C.~S.}\ \bibnamefont
  {Fischer}}, \bibinfo {author} {\bibfnamefont {A.}~\bibnamefont {Maas}},\ and\
  \bibinfo {author} {\bibfnamefont {J.~M.}\ \bibnamefont {Pawlowski}},\
  }\bibfield  {title} {\bibinfo {title} {{On the infrared behavior of Landau
  gauge Yang-Mills theory}},\ }\href
  {https://doi.org/10.1016/j.aop.2009.07.009} {\bibfield  {journal} {\bibinfo
  {journal} {Annals Phys.}\ }\textbf {\bibinfo {volume} {324}},\ \bibinfo
  {pages} {2408} (\bibinfo {year} {2009})},\ \Eprint
  {https://arxiv.org/abs/0810.1987} {arXiv:0810.1987 [hep-ph]} \BibitemShut
  {NoStop}%
\bibitem [{\citenamefont {Aguilar}\ \emph {et~al.}(2008)\citenamefont
  {Aguilar}, \citenamefont {Binosi},\ and\ \citenamefont
  {Papavassiliou}}]{Aguilar:2008xm}%
  \BibitemOpen
  \bibfield  {author} {\bibinfo {author} {\bibfnamefont {A.~C.}\ \bibnamefont
  {Aguilar}}, \bibinfo {author} {\bibfnamefont {D.}~\bibnamefont {Binosi}},\
  and\ \bibinfo {author} {\bibfnamefont {J.}~\bibnamefont {Papavassiliou}},\
  }\bibfield  {title} {\bibinfo {title} {{Gluon and ghost propagators in the
  Landau gauge: Deriving lattice results from Schwinger-Dyson equations}},\
  }\href {https://doi.org/10.1103/PhysRevD.78.025010} {\bibfield  {journal}
  {\bibinfo  {journal} {Phys. Rev. D}\ }\textbf {\bibinfo {volume} {78}},\
  \bibinfo {pages} {025010} (\bibinfo {year} {2008})},\ \Eprint
  {https://arxiv.org/abs/0802.1870} {arXiv:0802.1870 [hep-ph]} \BibitemShut
  {NoStop}%
\bibitem [{\citenamefont {Dudal}\ \emph
  {et~al.}(2008{\natexlab{a}})\citenamefont {Dudal}, \citenamefont {Sorella},
  \citenamefont {Vandersickel},\ and\ \citenamefont
  {Verschelde}}]{Dudal:2007cw}%
  \BibitemOpen
  \bibfield  {author} {\bibinfo {author} {\bibfnamefont {D.}~\bibnamefont
  {Dudal}}, \bibinfo {author} {\bibfnamefont {S.~P.}\ \bibnamefont {Sorella}},
  \bibinfo {author} {\bibfnamefont {N.}~\bibnamefont {Vandersickel}},\ and\
  \bibinfo {author} {\bibfnamefont {H.}~\bibnamefont {Verschelde}},\ }\bibfield
   {title} {\bibinfo {title} {{New features of the gluon and ghost propagator
  in the infrared region from the Gribov-Zwanziger approach}},\ }\href
  {https://doi.org/10.1103/PhysRevD.77.071501} {\bibfield  {journal} {\bibinfo
  {journal} {Phys. Rev.}\ }\textbf {\bibinfo {volume} {D77}},\ \bibinfo {pages}
  {071501} (\bibinfo {year} {2008}{\natexlab{a}})},\ \Eprint
  {https://arxiv.org/abs/0711.4496} {arXiv:0711.4496 [hep-th]} \BibitemShut
  {NoStop}%
%%CITATION = ARXIV:0711.4496;%%
\bibitem [{\citenamefont {Dudal}\ \emph
  {et~al.}(2008{\natexlab{b}})\citenamefont {Dudal}, \citenamefont {Gracey},
  \citenamefont {Sorella}, \citenamefont {Vandersickel},\ and\ \citenamefont
  {Verschelde}}]{Dudal:2008sp}%
  \BibitemOpen
  \bibfield  {author} {\bibinfo {author} {\bibfnamefont {D.}~\bibnamefont
  {Dudal}}, \bibinfo {author} {\bibfnamefont {J.~A.}\ \bibnamefont {Gracey}},
  \bibinfo {author} {\bibfnamefont {S.~P.}\ \bibnamefont {Sorella}}, \bibinfo
  {author} {\bibfnamefont {N.}~\bibnamefont {Vandersickel}},\ and\ \bibinfo
  {author} {\bibfnamefont {H.}~\bibnamefont {Verschelde}},\ }\bibfield  {title}
  {\bibinfo {title} {{A Refinement of the Gribov-Zwanziger approach in the
  Landau gauge: Infrared propagators in harmony with the lattice results}},\
  }\href {https://doi.org/10.1103/PhysRevD.78.065047} {\bibfield  {journal}
  {\bibinfo  {journal} {Phys. Rev.}\ }\textbf {\bibinfo {volume} {D78}},\
  \bibinfo {pages} {065047} (\bibinfo {year} {2008}{\natexlab{b}})},\ \Eprint
  {https://arxiv.org/abs/0806.4348} {arXiv:0806.4348 [hep-th]} \BibitemShut
  {NoStop}%
%%CITATION = ARXIV:0806.4348;%%
\bibitem [{\citenamefont {Pel\'aez}\ \emph {et~al.}(2021)\citenamefont
  {Pel\'aez}, \citenamefont {Reinosa}, \citenamefont {Serreau}, \citenamefont
  {Tissier},\ and\ \citenamefont {Wschebor}}]{Pelaez:2021tpq}%
  \BibitemOpen
  \bibfield  {author} {\bibinfo {author} {\bibfnamefont {M.}~\bibnamefont
  {Pel\'aez}}, \bibinfo {author} {\bibfnamefont {U.}~\bibnamefont {Reinosa}},
  \bibinfo {author} {\bibfnamefont {J.}~\bibnamefont {Serreau}}, \bibinfo
  {author} {\bibfnamefont {M.}~\bibnamefont {Tissier}},\ and\ \bibinfo {author}
  {\bibfnamefont {N.}~\bibnamefont {Wschebor}},\ }\bibfield  {title} {\bibinfo
  {title} {{A window on infrared QCD with small expansion parameters}},\ }\href
  {https://doi.org/10.1088/1361-6633/ac36b8} {\bibfield  {journal} {\bibinfo
  {journal} {Rept. Prog. Phys.}\ }\textbf {\bibinfo {volume} {84}},\ \bibinfo
  {pages} {124202} (\bibinfo {year} {2021})},\ \Eprint
  {https://arxiv.org/abs/2106.04526} {arXiv:2106.04526 [hep-th]} \BibitemShut
  {NoStop}%
\bibitem [{\citenamefont {Pel\'aez}\ \emph {et~al.}(2017)\citenamefont
  {Pel\'aez}, \citenamefont {Reinosa}, \citenamefont {Serreau}, \citenamefont
  {Tissier},\ and\ \citenamefont {Wschebor}}]{Pelaez:2017bhh}%
  \BibitemOpen
  \bibfield  {author} {\bibinfo {author} {\bibfnamefont {M.}~\bibnamefont
  {Pel\'aez}}, \bibinfo {author} {\bibfnamefont {U.}~\bibnamefont {Reinosa}},
  \bibinfo {author} {\bibfnamefont {J.}~\bibnamefont {Serreau}}, \bibinfo
  {author} {\bibfnamefont {M.}~\bibnamefont {Tissier}},\ and\ \bibinfo {author}
  {\bibfnamefont {N.}~\bibnamefont {Wschebor}},\ }\bibfield  {title} {\bibinfo
  {title} {{Small parameters in infrared quantum chromodynamics}},\ }\href
  {https://doi.org/10.1103/PhysRevD.96.114011} {\bibfield  {journal} {\bibinfo
  {journal} {Phys. Rev. D}\ }\textbf {\bibinfo {volume} {96}},\ \bibinfo
  {pages} {114011} (\bibinfo {year} {2017})},\ \Eprint
  {https://arxiv.org/abs/1703.10288} {arXiv:1703.10288 [hep-th]} \BibitemShut
  {NoStop}%
\bibitem [{\citenamefont {Gribov}(1978)}]{Gribov:1977wm}%
  \BibitemOpen
  \bibfield  {author} {\bibinfo {author} {\bibfnamefont {V.~N.}\ \bibnamefont
  {Gribov}},\ }\bibfield  {title} {\bibinfo {title} {{Quantization of
  Nonabelian Gauge Theories}},\ }\href
  {https://doi.org/10.1016/0550-3213(78)90175-X} {\bibfield  {journal}
  {\bibinfo  {journal} {Nucl. Phys.}\ }\textbf {\bibinfo {volume} {B139}},\
  \bibinfo {pages} {1} (\bibinfo {year} {1978})},\ \bibinfo {note}
  {[1(1977)]}\BibitemShut {NoStop}%
%%CITATION = NUPHA,B139,1;%%
\bibitem [{\citenamefont {Singer}(1978)}]{Singer:1978dk}%
  \BibitemOpen
  \bibfield  {author} {\bibinfo {author} {\bibfnamefont {I.}~\bibnamefont
  {Singer}},\ }\bibfield  {title} {\bibinfo {title} {{Some Remarks on the
  Gribov Ambiguity}},\ }\href {https://doi.org/10.1007/BF01609471} {\bibfield
  {journal} {\bibinfo  {journal} {Commun. Math. Phys.}\ }\textbf {\bibinfo
  {volume} {60}},\ \bibinfo {pages} {7} (\bibinfo {year} {1978})}\BibitemShut
  {NoStop}%
\bibitem [{\citenamefont {Sobreiro}\ and\ \citenamefont
  {Sorella}(2005)}]{Sobreiro:2005ec}%
  \BibitemOpen
  \bibfield  {author} {\bibinfo {author} {\bibfnamefont {R.}~\bibnamefont
  {Sobreiro}}\ and\ \bibinfo {author} {\bibfnamefont {S.}~\bibnamefont
  {Sorella}},\ }\bibfield  {title} {\bibinfo {title} {{Introduction to the
  Gribov ambiguities in Euclidean Yang-Mills theories}},\ }in\ \href@noop {}
  {\emph {\bibinfo {booktitle} {{13th Jorge Andre Swieca Summer School on
  Particle and Fields}}}}\ (\bibinfo {year} {2005})\ \Eprint
  {https://arxiv.org/abs/hep-th/0504095} {arXiv:hep-th/0504095} \BibitemShut
  {NoStop}%
\bibitem [{\citenamefont {Vandersickel}(2011)}]{Vandersickel:2011zc}%
  \BibitemOpen
  \bibfield  {author} {\bibinfo {author} {\bibfnamefont {N.}~\bibnamefont
  {Vandersickel}},\ }\emph {\bibinfo {title} {{A Study of the Gribov-Zwanziger
  action: from propagators to glueballs}}},\ \href@noop {} {\bibinfo {type}
  {Other thesis}} (\bibinfo {year} {2011}),\ \Eprint
  {https://arxiv.org/abs/1104.1315} {arXiv:1104.1315 [hep-th]} \BibitemShut
  {NoStop}%
\bibitem [{\citenamefont {Vandersickel}\ and\ \citenamefont
  {Zwanziger}(2012)}]{Vandersickel:2012tz}%
  \BibitemOpen
  \bibfield  {author} {\bibinfo {author} {\bibfnamefont {N.}~\bibnamefont
  {Vandersickel}}\ and\ \bibinfo {author} {\bibfnamefont {D.}~\bibnamefont
  {Zwanziger}},\ }\bibfield  {title} {\bibinfo {title} {{The Gribov problem and
  QCD dynamics}},\ }\href {https://doi.org/10.1016/j.physrep.2012.07.003}
  {\bibfield  {journal} {\bibinfo  {journal} {Phys. Rept.}\ }\textbf {\bibinfo
  {volume} {520}},\ \bibinfo {pages} {175} (\bibinfo {year} {2012})},\ \Eprint
  {https://arxiv.org/abs/1202.1491} {arXiv:1202.1491 [hep-th]} \BibitemShut
  {NoStop}%
\bibitem [{\citenamefont {Zwanziger}(1989)}]{Zwanziger:1989mf}%
  \BibitemOpen
  \bibfield  {author} {\bibinfo {author} {\bibfnamefont {D.}~\bibnamefont
  {Zwanziger}},\ }\bibfield  {title} {\bibinfo {title} {{Local and
  Renormalizable Action From the Gribov Horizon}},\ }\href
  {https://doi.org/10.1016/0550-3213(89)90122-3} {\bibfield  {journal}
  {\bibinfo  {journal} {Nucl. Phys.}\ }\textbf {\bibinfo {volume} {B323}},\
  \bibinfo {pages} {513} (\bibinfo {year} {1989})}\BibitemShut {NoStop}%
%%CITATION = NUPHA,B323,513;%%
\bibitem [{\citenamefont {Capri}\ \emph {et~al.}(2005)\citenamefont {Capri},
  \citenamefont {Lemes}, \citenamefont {Sobreiro}, \citenamefont {Sorella},\
  and\ \citenamefont {Thibes}}]{Capri:2005tj}%
  \BibitemOpen
  \bibfield  {author} {\bibinfo {author} {\bibfnamefont {M.~A.~L.}\
  \bibnamefont {Capri}}, \bibinfo {author} {\bibfnamefont {V.~E.~R.}\
  \bibnamefont {Lemes}}, \bibinfo {author} {\bibfnamefont {R.~F.}\ \bibnamefont
  {Sobreiro}}, \bibinfo {author} {\bibfnamefont {S.~P.}\ \bibnamefont
  {Sorella}},\ and\ \bibinfo {author} {\bibfnamefont {R.}~\bibnamefont
  {Thibes}},\ }\bibfield  {title} {\bibinfo {title} {{The Influence of the
  Gribov copies on the gluon and ghost propagators in Euclidean Yang-Mills
  theory in the maximal Abelian gauge}},\ }\href
  {https://doi.org/10.1103/PhysRevD.72.085021} {\bibfield  {journal} {\bibinfo
  {journal} {Phys. Rev. D}\ }\textbf {\bibinfo {volume} {72}},\ \bibinfo
  {pages} {085021} (\bibinfo {year} {2005})},\ \Eprint
  {https://arxiv.org/abs/hep-th/0507052} {arXiv:hep-th/0507052} \BibitemShut
  {NoStop}%
\bibitem [{\citenamefont {Gongyo}\ and\ \citenamefont
  {Iida}(2014)}]{Gongyo:2013rua}%
  \BibitemOpen
  \bibfield  {author} {\bibinfo {author} {\bibfnamefont {S.}~\bibnamefont
  {Gongyo}}\ and\ \bibinfo {author} {\bibfnamefont {H.}~\bibnamefont {Iida}},\
  }\bibfield  {title} {\bibinfo {title} {{Gribov-Zwanziger action in SU(2)
  maximally Abelian gauge with U(1)$_3$ Landau gauge}},\ }\href
  {https://doi.org/10.1103/PhysRevD.89.025022} {\bibfield  {journal} {\bibinfo
  {journal} {Phys. Rev. D}\ }\textbf {\bibinfo {volume} {89}},\ \bibinfo
  {pages} {025022} (\bibinfo {year} {2014})},\ \Eprint
  {https://arxiv.org/abs/1310.4877} {arXiv:1310.4877 [hep-th]} \BibitemShut
  {NoStop}%
\bibitem [{\citenamefont {Nambu}(1974)}]{Nambu1974}%
  \BibitemOpen
  \bibfield  {author} {\bibinfo {author} {\bibfnamefont {Y.}~\bibnamefont
  {Nambu}},\ }\bibfield  {title} {\bibinfo {title} {Strings, monopoles, and
  gauge fields},\ }\href {https://doi.org/10.1103/PhysRevD.10.4262} {\bibfield
  {journal} {\bibinfo  {journal} {Phys. Rev. D}\ }\textbf {\bibinfo {volume}
  {10}},\ \bibinfo {pages} {4262} (\bibinfo {year} {1974})}\BibitemShut
  {NoStop}%
\bibitem [{\citenamefont {’t Hooft}(1975)}]{tHooft1975}%
  \BibitemOpen
  \bibfield  {author} {\bibinfo {author} {\bibfnamefont {G.}~\bibnamefont {’t
  Hooft}},\ }\bibfield  {title} {\bibinfo {title} {Gauge fields with unified
  weak, electromagnetic and strong interactions},\ }in\ \href@noop {} {\emph
  {\bibinfo {booktitle} {High Energy Physics}}},\ \bibinfo {editor} {edited by\
  \bibinfo {editor} {\bibfnamefont {A.}~\bibnamefont {Zichichi}}}\ (\bibinfo
  {address} {Palermo},\ \bibinfo {year} {1975})\BibitemShut {NoStop}%
\bibitem [{\citenamefont {Mandelstam}(1976)}]{Mandelstam1976}%
  \BibitemOpen
  \bibfield  {author} {\bibinfo {author} {\bibfnamefont {S.}~\bibnamefont
  {Mandelstam}},\ }\bibfield  {title} {\bibinfo {title} {Vortices and quark
  confinement in non-abelian gauge theories},\ }\href
  {https://doi.org/10.1016/0370-1573(76)90043-0} {\bibfield  {journal}
  {\bibinfo  {journal} {Phys. Rept.}\ }\textbf {\bibinfo {volume} {23}},\
  \bibinfo {pages} {245} (\bibinfo {year} {1976})}\BibitemShut {NoStop}%
\bibitem [{\citenamefont {Ezawa}\ and\ \citenamefont
  {Iwazaki}(1982{\natexlab{a}})}]{Ezawa1982}%
  \BibitemOpen
  \bibfield  {author} {\bibinfo {author} {\bibfnamefont {Z.~F.}\ \bibnamefont
  {Ezawa}}\ and\ \bibinfo {author} {\bibfnamefont {A.}~\bibnamefont
  {Iwazaki}},\ }\bibfield  {title} {\bibinfo {title} {Abelian dominance and
  quark confinement in yang--mills theories},\ }\href
  {https://doi.org/10.1103/PhysRevD.25.2681} {\bibfield  {journal} {\bibinfo
  {journal} {Phys. Rev. D}\ }\textbf {\bibinfo {volume} {25}},\ \bibinfo
  {pages} {2681} (\bibinfo {year} {1982}{\natexlab{a}})}\BibitemShut {NoStop}%
\bibitem [{\citenamefont {Suzuki}\ and\ \citenamefont
  {Yotsuyanagi}(1990{\natexlab{a}})}]{Suzuki1990}%
  \BibitemOpen
  \bibfield  {author} {\bibinfo {author} {\bibfnamefont {T.}~\bibnamefont
  {Suzuki}}\ and\ \bibinfo {author} {\bibfnamefont {I.}~\bibnamefont
  {Yotsuyanagi}},\ }\bibfield  {title} {\bibinfo {title} {Possible evidence of
  abelian dominance in quark confinement},\ }\href
  {https://doi.org/10.1103/PhysRevD.42.4257} {\bibfield  {journal} {\bibinfo
  {journal} {Phys. Rev. D}\ }\textbf {\bibinfo {volume} {42}},\ \bibinfo
  {pages} {4257} (\bibinfo {year} {1990}{\natexlab{a}})}\BibitemShut {NoStop}%
\bibitem [{\citenamefont {Suzuki}\ \emph {et~al.}(1992)\citenamefont {Suzuki},
  \citenamefont {Hioki}, \citenamefont {Kitahara}, \citenamefont {Kiura},
  \citenamefont {Matsubara}, \citenamefont {Miyamaura},\ and\ \citenamefont
  {Ohno}}]{Suzuki:1992gz}%
  \BibitemOpen
  \bibfield  {author} {\bibinfo {author} {\bibfnamefont {T.}~\bibnamefont
  {Suzuki}}, \bibinfo {author} {\bibfnamefont {S.}~\bibnamefont {Hioki}},
  \bibinfo {author} {\bibfnamefont {S.}~\bibnamefont {Kitahara}}, \bibinfo
  {author} {\bibfnamefont {S.}~\bibnamefont {Kiura}}, \bibinfo {author}
  {\bibfnamefont {Y.}~\bibnamefont {Matsubara}}, \bibinfo {author}
  {\bibfnamefont {O.}~\bibnamefont {Miyamaura}},\ and\ \bibinfo {author}
  {\bibfnamefont {S.}~\bibnamefont {Ohno}},\ }\bibfield  {title} {\bibinfo
  {title} {{Abelian dominance in SU(2) color confinement}},\ }\href
  {https://doi.org/10.1016/0920-5632(92)90298-7} {\bibfield  {journal}
  {\bibinfo  {journal} {Nucl. Phys. B Proc. Suppl.}\ }\textbf {\bibinfo
  {volume} {26}},\ \bibinfo {pages} {441} (\bibinfo {year} {1992})}\BibitemShut
  {NoStop}%
\bibitem [{\citenamefont {Hioki}\ \emph {et~al.}(1991)\citenamefont {Hioki},
  \citenamefont {Kitahara}, \citenamefont {Kiura}, \citenamefont {Matsubara},
  \citenamefont {Miyamura}, \citenamefont {Ohno},\ and\ \citenamefont
  {Suzuki}}]{Hioki1991}%
  \BibitemOpen
  \bibfield  {author} {\bibinfo {author} {\bibfnamefont {S.}~\bibnamefont
  {Hioki}}, \bibinfo {author} {\bibfnamefont {S.}~\bibnamefont {Kitahara}},
  \bibinfo {author} {\bibfnamefont {S.}~\bibnamefont {Kiura}}, \bibinfo
  {author} {\bibfnamefont {Y.}~\bibnamefont {Matsubara}}, \bibinfo {author}
  {\bibfnamefont {O.}~\bibnamefont {Miyamura}}, \bibinfo {author}
  {\bibfnamefont {S.}~\bibnamefont {Ohno}},\ and\ \bibinfo {author}
  {\bibfnamefont {T.}~\bibnamefont {Suzuki}},\ }\bibfield  {title} {\bibinfo
  {title} {Abelian dominance in su(2) lattice gauge theory},\ }\href
  {https://doi.org/10.1016/0370-2693(91)91802-2} {\bibfield  {journal}
  {\bibinfo  {journal} {Phys. Lett. B}\ }\textbf {\bibinfo {volume} {272}},\
  \bibinfo {pages} {326} (\bibinfo {year} {1991})},\ \bibinfo {note} {erratum:
  Phys. Lett. B 281 (1992) 416}\BibitemShut {NoStop}%
\bibitem [{\citenamefont {Sakumichi}\ and\ \citenamefont
  {Suganuma}(2014)}]{Sakumichi:2014xpa}%
  \BibitemOpen
  \bibfield  {author} {\bibinfo {author} {\bibfnamefont {N.}~\bibnamefont
  {Sakumichi}}\ and\ \bibinfo {author} {\bibfnamefont {H.}~\bibnamefont
  {Suganuma}},\ }\bibfield  {title} {\bibinfo {title} {{Perfect Abelian
  dominance of quark confinement in SU(3) QCD}},\ }\href
  {https://doi.org/10.1103/PhysRevD.90.111501} {\bibfield  {journal} {\bibinfo
  {journal} {Phys. Rev. D}\ }\textbf {\bibinfo {volume} {90}},\ \bibinfo
  {pages} {111501} (\bibinfo {year} {2014})},\ \Eprint
  {https://arxiv.org/abs/1406.2215} {arXiv:1406.2215 [hep-lat]} \BibitemShut
  {NoStop}%
\bibitem [{\citenamefont {Capri}\ \emph {et~al.}(2008)\citenamefont {Capri},
  \citenamefont {Lemes}, \citenamefont {Sobreiro}, \citenamefont {Sorella},\
  and\ \citenamefont {Thibes}}]{Capri:2008ak}%
  \BibitemOpen
  \bibfield  {author} {\bibinfo {author} {\bibfnamefont {M.~A.~L.}\
  \bibnamefont {Capri}}, \bibinfo {author} {\bibfnamefont {V.~E.~R.}\
  \bibnamefont {Lemes}}, \bibinfo {author} {\bibfnamefont {R.~F.}\ \bibnamefont
  {Sobreiro}}, \bibinfo {author} {\bibfnamefont {S.~P.}\ \bibnamefont
  {Sorella}},\ and\ \bibinfo {author} {\bibfnamefont {R.}~\bibnamefont
  {Thibes}},\ }\bibfield  {title} {\bibinfo {title} {{The Gluon and ghost
  propagators in Euclidean Yang-Mills theory in the maximal Abelian gauge:
  Taking into account the effects of the Gribov copies and of the dimension two
  condensates}},\ }\href {https://doi.org/10.1103/PhysRevD.77.105023}
  {\bibfield  {journal} {\bibinfo  {journal} {Phys. Rev. D}\ }\textbf {\bibinfo
  {volume} {77}},\ \bibinfo {pages} {105023} (\bibinfo {year} {2008})},\
  \Eprint {https://arxiv.org/abs/0801.0566} {arXiv:0801.0566 [hep-th]}
  \BibitemShut {NoStop}%
\bibitem [{\citenamefont {Capri}\ \emph {et~al.}(2009)\citenamefont {Capri},
  \citenamefont {Gomez}, \citenamefont {Lemes}, \citenamefont {Sobreiro},\ and\
  \citenamefont {Sorella}}]{Capri:2008vk}%
  \BibitemOpen
  \bibfield  {author} {\bibinfo {author} {\bibfnamefont {M.~A.~L.}\
  \bibnamefont {Capri}}, \bibinfo {author} {\bibfnamefont {A.~J.}\ \bibnamefont
  {Gomez}}, \bibinfo {author} {\bibfnamefont {V.~E.~R.}\ \bibnamefont {Lemes}},
  \bibinfo {author} {\bibfnamefont {R.~F.}\ \bibnamefont {Sobreiro}},\ and\
  \bibinfo {author} {\bibfnamefont {S.~P.}\ \bibnamefont {Sorella}},\
  }\bibfield  {title} {\bibinfo {title} {{Study of the Gribov region in
  Euclidean Yang-Mills theories in the maximal Abelian gauge}},\ }\href
  {https://doi.org/10.1103/PhysRevD.79.025019} {\bibfield  {journal} {\bibinfo
  {journal} {Phys. Rev. D}\ }\textbf {\bibinfo {volume} {79}},\ \bibinfo
  {pages} {025019} (\bibinfo {year} {2009})},\ \Eprint
  {https://arxiv.org/abs/0811.2760} {arXiv:0811.2760 [hep-th]} \BibitemShut
  {NoStop}%
\bibitem [{\citenamefont {Capri}\ \emph
  {et~al.}(2015{\natexlab{a}})\citenamefont {Capri}, \citenamefont {Dudal},
  \citenamefont {Fiorentini}, \citenamefont {Guimaraes}, \citenamefont {Justo},
  \citenamefont {Pereira}, \citenamefont {Mintz}, \citenamefont {Palhares},
  \citenamefont {Sobreiro},\ and\ \citenamefont {Sorella}}]{Capri:2015ixa}%
  \BibitemOpen
  \bibfield  {author} {\bibinfo {author} {\bibfnamefont {M.~A.~L.}\
  \bibnamefont {Capri}}, \bibinfo {author} {\bibfnamefont {D.}~\bibnamefont
  {Dudal}}, \bibinfo {author} {\bibfnamefont {D.}~\bibnamefont {Fiorentini}},
  \bibinfo {author} {\bibfnamefont {M.~S.}\ \bibnamefont {Guimaraes}}, \bibinfo
  {author} {\bibfnamefont {I.~F.}\ \bibnamefont {Justo}}, \bibinfo {author}
  {\bibfnamefont {A.~D.}\ \bibnamefont {Pereira}}, \bibinfo {author}
  {\bibfnamefont {B.~W.}\ \bibnamefont {Mintz}}, \bibinfo {author}
  {\bibfnamefont {L.~F.}\ \bibnamefont {Palhares}}, \bibinfo {author}
  {\bibfnamefont {R.~F.}\ \bibnamefont {Sobreiro}},\ and\ \bibinfo {author}
  {\bibfnamefont {S.~P.}\ \bibnamefont {Sorella}},\ }\bibfield  {title}
  {\bibinfo {title} {{Exact nilpotent nonperturbative BRST symmetry for the
  Gribov-Zwanziger action in the linear covariant gauge}},\ }\href
  {https://doi.org/10.1103/PhysRevD.92.045039} {\bibfield  {journal} {\bibinfo
  {journal} {Phys. Rev. D}\ }\textbf {\bibinfo {volume} {92}},\ \bibinfo
  {pages} {045039} (\bibinfo {year} {2015}{\natexlab{a}})},\ \Eprint
  {https://arxiv.org/abs/1506.06995} {arXiv:1506.06995 [hep-th]} \BibitemShut
  {NoStop}%
\bibitem [{\citenamefont {Capri}\ \emph
  {et~al.}(2016{\natexlab{a}})\citenamefont {Capri}, \citenamefont
  {Fiorentini}, \citenamefont {Guimaraes}, \citenamefont {Mintz}, \citenamefont
  {Palhares}, \citenamefont {Sorella}, \citenamefont {Dudal}, \citenamefont
  {Justo}, \citenamefont {Pereira},\ and\ \citenamefont
  {Sobreiro}}]{Capri:2015nzw}%
  \BibitemOpen
  \bibfield  {author} {\bibinfo {author} {\bibfnamefont {M.~A.~L.}\
  \bibnamefont {Capri}}, \bibinfo {author} {\bibfnamefont {D.}~\bibnamefont
  {Fiorentini}}, \bibinfo {author} {\bibfnamefont {M.~S.}\ \bibnamefont
  {Guimaraes}}, \bibinfo {author} {\bibfnamefont {B.~W.}\ \bibnamefont
  {Mintz}}, \bibinfo {author} {\bibfnamefont {L.~F.}\ \bibnamefont {Palhares}},
  \bibinfo {author} {\bibfnamefont {S.~P.}\ \bibnamefont {Sorella}}, \bibinfo
  {author} {\bibfnamefont {D.}~\bibnamefont {Dudal}}, \bibinfo {author}
  {\bibfnamefont {I.~F.}\ \bibnamefont {Justo}}, \bibinfo {author}
  {\bibfnamefont {A.~D.}\ \bibnamefont {Pereira}},\ and\ \bibinfo {author}
  {\bibfnamefont {R.~F.}\ \bibnamefont {Sobreiro}},\ }\bibfield  {title}
  {\bibinfo {title} {{More on the nonperturbative Gribov-Zwanziger quantization
  of linear covariant gauges}},\ }\href
  {https://doi.org/10.1103/PhysRevD.93.065019} {\bibfield  {journal} {\bibinfo
  {journal} {Phys. Rev. D}\ }\textbf {\bibinfo {volume} {93}},\ \bibinfo
  {pages} {065019} (\bibinfo {year} {2016}{\natexlab{a}})},\ \Eprint
  {https://arxiv.org/abs/1512.05833} {arXiv:1512.05833 [hep-th]} \BibitemShut
  {NoStop}%
\bibitem [{\citenamefont {Capri}\ \emph
  {et~al.}(2017{\natexlab{a}})\citenamefont {Capri}, \citenamefont
  {Fiorentini}, \citenamefont {Pereira}, \citenamefont {Sobreiro},
  \citenamefont {Sorella},\ and\ \citenamefont {Terin}}]{Capri:2016aif}%
  \BibitemOpen
  \bibfield  {author} {\bibinfo {author} {\bibfnamefont {M.~A.~L.}\
  \bibnamefont {Capri}}, \bibinfo {author} {\bibfnamefont {D.}~\bibnamefont
  {Fiorentini}}, \bibinfo {author} {\bibfnamefont {A.~D.}\ \bibnamefont
  {Pereira}}, \bibinfo {author} {\bibfnamefont {R.~F.}\ \bibnamefont
  {Sobreiro}}, \bibinfo {author} {\bibfnamefont {S.~P.}\ \bibnamefont
  {Sorella}},\ and\ \bibinfo {author} {\bibfnamefont {R.~C.}\ \bibnamefont
  {Terin}},\ }\bibfield  {title} {\bibinfo {title} {{Aspects of the refined
  Gribov-Zwanziger action in linear covariant gauges}},\ }\href
  {https://doi.org/10.1016/j.aop.2016.10.023} {\bibfield  {journal} {\bibinfo
  {journal} {Annals Phys.}\ }\textbf {\bibinfo {volume} {376}},\ \bibinfo
  {pages} {40} (\bibinfo {year} {2017}{\natexlab{a}})},\ \Eprint
  {https://arxiv.org/abs/1607.07912} {arXiv:1607.07912 [hep-th]} \BibitemShut
  {NoStop}%
\bibitem [{\citenamefont {Capri}\ \emph
  {et~al.}(2016{\natexlab{b}})\citenamefont {Capri}, \citenamefont
  {Fiorentini}, \citenamefont {Guimaraes}, \citenamefont {Mintz}, \citenamefont
  {Palhares},\ and\ \citenamefont {Sorella}}]{Capri:2016ovw}%
  \BibitemOpen
  \bibfield  {author} {\bibinfo {author} {\bibfnamefont {M.~A.~L.}\
  \bibnamefont {Capri}}, \bibinfo {author} {\bibfnamefont {D.}~\bibnamefont
  {Fiorentini}}, \bibinfo {author} {\bibfnamefont {M.~S.}\ \bibnamefont
  {Guimaraes}}, \bibinfo {author} {\bibfnamefont {B.~W.}\ \bibnamefont
  {Mintz}}, \bibinfo {author} {\bibfnamefont {L.~F.}\ \bibnamefont
  {Palhares}},\ and\ \bibinfo {author} {\bibfnamefont {S.~P.}\ \bibnamefont
  {Sorella}},\ }\bibfield  {title} {\bibinfo {title} {{Local and renormalizable
  framework for the gauge-invariant operator $A^2_{\min}$ in Euclidean
  Yang-Mills theories in linear covariant gauges}},\ }\href
  {https://doi.org/10.1103/PhysRevD.94.065009} {\bibfield  {journal} {\bibinfo
  {journal} {Phys. Rev. D}\ }\textbf {\bibinfo {volume} {94}},\ \bibinfo
  {pages} {065009} (\bibinfo {year} {2016}{\natexlab{b}})},\ \Eprint
  {https://arxiv.org/abs/1606.06601} {arXiv:1606.06601 [hep-th]} \BibitemShut
  {NoStop}%
\bibitem [{\citenamefont {Capri}\ \emph
  {et~al.}(2017{\natexlab{b}})\citenamefont {Capri}, \citenamefont
  {Fiorentini}, \citenamefont {Pereira},\ and\ \citenamefont
  {Sorella}}]{Capri:2017bfd}%
  \BibitemOpen
  \bibfield  {author} {\bibinfo {author} {\bibfnamefont {M.~A.~L.}\
  \bibnamefont {Capri}}, \bibinfo {author} {\bibfnamefont {D.}~\bibnamefont
  {Fiorentini}}, \bibinfo {author} {\bibfnamefont {A.~D.}\ \bibnamefont
  {Pereira}},\ and\ \bibinfo {author} {\bibfnamefont {S.~P.}\ \bibnamefont
  {Sorella}},\ }\bibfield  {title} {\bibinfo {title} {{Renormalizability of the
  refined Gribov-Zwanziger action in linear covariant gauges}},\ }\href
  {https://doi.org/10.1103/PhysRevD.96.054022} {\bibfield  {journal} {\bibinfo
  {journal} {Phys. Rev. D}\ }\textbf {\bibinfo {volume} {96}},\ \bibinfo
  {pages} {054022} (\bibinfo {year} {2017}{\natexlab{b}})},\ \Eprint
  {https://arxiv.org/abs/1708.01543} {arXiv:1708.01543 [hep-th]} \BibitemShut
  {NoStop}%
\bibitem [{\citenamefont {Pereira}\ \emph {et~al.}(2016)\citenamefont
  {Pereira}, \citenamefont {Sobreiro},\ and\ \citenamefont
  {Sorella}}]{Pereira:2016fpn}%
  \BibitemOpen
  \bibfield  {author} {\bibinfo {author} {\bibfnamefont {A.~D.}\ \bibnamefont
  {Pereira}}, \bibinfo {author} {\bibfnamefont {R.~F.}\ \bibnamefont
  {Sobreiro}},\ and\ \bibinfo {author} {\bibfnamefont {S.~P.}\ \bibnamefont
  {Sorella}},\ }\bibfield  {title} {\bibinfo {title} {{Non-perturbative BRST
  quantization of Euclidean Yang\textendash{}Mills theories in
  Curci\textendash{}Ferrari gauges}},\ }\href
  {https://doi.org/10.1140/epjc/s10052-016-4368-2} {\bibfield  {journal}
  {\bibinfo  {journal} {Eur. Phys. J. C}\ }\textbf {\bibinfo {volume} {76}},\
  \bibinfo {pages} {528} (\bibinfo {year} {2016})},\ \Eprint
  {https://arxiv.org/abs/1605.09747} {arXiv:1605.09747 [hep-th]} \BibitemShut
  {NoStop}%
\bibitem [{\citenamefont {Serreau}\ and\ \citenamefont
  {Tissier}(2012)}]{Serreau:2012cg}%
  \BibitemOpen
  \bibfield  {author} {\bibinfo {author} {\bibfnamefont {J.}~\bibnamefont
  {Serreau}}\ and\ \bibinfo {author} {\bibfnamefont {M.}~\bibnamefont
  {Tissier}},\ }\bibfield  {title} {\bibinfo {title} {{Lifting the Gribov
  ambiguity in Yang-Mills theories}},\ }\href
  {https://doi.org/10.1016/j.physletb.2012.04.041} {\bibfield  {journal}
  {\bibinfo  {journal} {Phys. Lett. B}\ }\textbf {\bibinfo {volume} {712}},\
  \bibinfo {pages} {97} (\bibinfo {year} {2012})},\ \Eprint
  {https://arxiv.org/abs/1202.3432} {arXiv:1202.3432 [hep-th]} \BibitemShut
  {NoStop}%
\bibitem [{\citenamefont {Serreau}\ \emph {et~al.}(2014)\citenamefont
  {Serreau}, \citenamefont {Tissier},\ and\ \citenamefont
  {Tresmontant}}]{Serreau:2013ila}%
  \BibitemOpen
  \bibfield  {author} {\bibinfo {author} {\bibfnamefont {J.}~\bibnamefont
  {Serreau}}, \bibinfo {author} {\bibfnamefont {M.}~\bibnamefont {Tissier}},\
  and\ \bibinfo {author} {\bibfnamefont {A.}~\bibnamefont {Tresmontant}},\
  }\bibfield  {title} {\bibinfo {title} {{Covariant gauges without Gribov
  ambiguities in Yang-Mills theories}},\ }\href
  {https://doi.org/10.1103/PhysRevD.89.125019} {\bibfield  {journal} {\bibinfo
  {journal} {Phys. Rev. D}\ }\textbf {\bibinfo {volume} {89}},\ \bibinfo
  {pages} {125019} (\bibinfo {year} {2014})},\ \Eprint
  {https://arxiv.org/abs/1307.6019} {arXiv:1307.6019 [hep-th]} \BibitemShut
  {NoStop}%
\bibitem [{\citenamefont {Serreau}\ \emph {et~al.}(2015)\citenamefont
  {Serreau}, \citenamefont {Tissier},\ and\ \citenamefont
  {Tresmontant}}]{Serreau:2015yna}%
  \BibitemOpen
  \bibfield  {author} {\bibinfo {author} {\bibfnamefont {J.}~\bibnamefont
  {Serreau}}, \bibinfo {author} {\bibfnamefont {M.}~\bibnamefont {Tissier}},\
  and\ \bibinfo {author} {\bibfnamefont {A.}~\bibnamefont {Tresmontant}},\
  }\bibfield  {title} {\bibinfo {title} {{Influence of Gribov ambiguities in a
  class of nonlinear covariant gauges}},\ }\href
  {https://doi.org/10.1103/PhysRevD.92.105003} {\bibfield  {journal} {\bibinfo
  {journal} {Phys. Rev. D}\ }\textbf {\bibinfo {volume} {92}},\ \bibinfo
  {pages} {105003} (\bibinfo {year} {2015})},\ \Eprint
  {https://arxiv.org/abs/1505.07270} {arXiv:1505.07270 [hep-th]} \BibitemShut
  {NoStop}%
\bibitem [{\citenamefont {Reinosa}\ \emph {et~al.}(2021)\citenamefont
  {Reinosa}, \citenamefont {Serreau}, \citenamefont {Terin},\ and\
  \citenamefont {Tissier}}]{Reinosa:2020skx}%
  \BibitemOpen
  \bibfield  {author} {\bibinfo {author} {\bibfnamefont {U.}~\bibnamefont
  {Reinosa}}, \bibinfo {author} {\bibfnamefont {J.}~\bibnamefont {Serreau}},
  \bibinfo {author} {\bibfnamefont {R.~C.}\ \bibnamefont {Terin}},\ and\
  \bibinfo {author} {\bibfnamefont {M.}~\bibnamefont {Tissier}},\ }\bibfield
  {title} {\bibinfo {title} {{Symmetry restoration and the gluon mass in the
  Landau gauge}},\ }\href {https://doi.org/10.21468/SciPostPhys.10.2.035}
  {\bibfield  {journal} {\bibinfo  {journal} {SciPost Phys.}\ }\textbf
  {\bibinfo {volume} {10}},\ \bibinfo {pages} {035} (\bibinfo {year} {2021})},\
  \Eprint {https://arxiv.org/abs/2004.12413} {arXiv:2004.12413 [hep-th]}
  \BibitemShut {NoStop}%
\bibitem [{\citenamefont
  {Carmo~Terin}(2026{\natexlab{a}})}]{CarmoTerin:2025agt}%
  \BibitemOpen
  \bibfield  {author} {\bibinfo {author} {\bibfnamefont {R.}~\bibnamefont
  {Carmo~Terin}},\ }\bibfield  {title} {\bibinfo {title} {{Towards a unified
  viewpoint of Gribov{\textendash}Zwanziger and Serreau{\textendash}Tissier
  gauge fixing}},\ }\href {https://doi.org/10.1016/j.physletb.2026.140302}
  {\bibfield  {journal} {\bibinfo  {journal} {Phys. Lett. B}\ }\textbf
  {\bibinfo {volume} {875}},\ \bibinfo {pages} {140302} (\bibinfo {year}
  {2026}{\natexlab{a}})},\ \Eprint {https://arxiv.org/abs/2510.18443}
  {arXiv:2510.18443 [hep-th]} \BibitemShut {NoStop}%
\bibitem [{\citenamefont
  {Carmo~Terin}(2026{\natexlab{b}})}]{CarmoTerin:2026rzb}%
  \BibitemOpen
  \bibfield  {author} {\bibinfo {author} {\bibfnamefont {R.}~\bibnamefont
  {Carmo~Terin}},\ }\bibfield  {title} {\bibinfo {title} {{Emergent Gribov
  horizon from replica symmetry breaking in Yang--Mills theories}},\
  }\href@noop {} {\  (\bibinfo {year} {2026}{\natexlab{b}})},\ \Eprint
  {https://arxiv.org/abs/2603.02838} {arXiv:2603.02838 [hep-th]} \BibitemShut
  {NoStop}%
\bibitem [{\citenamefont {Eichmann}\ \emph {et~al.}(2021)\citenamefont
  {Eichmann}, \citenamefont {Pawlowski},\ and\ \citenamefont
  {Silva}}]{Eichmann:2021zuv}%
  \BibitemOpen
  \bibfield  {author} {\bibinfo {author} {\bibfnamefont {G.}~\bibnamefont
  {Eichmann}}, \bibinfo {author} {\bibfnamefont {J.~M.}\ \bibnamefont
  {Pawlowski}},\ and\ \bibinfo {author} {\bibfnamefont {J.~M.}\ \bibnamefont
  {Silva}},\ }\bibfield  {title} {\bibinfo {title} {{Mass generation in
  Landau-gauge Yang-Mills theory}},\ }\href
  {https://doi.org/10.1103/PhysRevD.104.114016} {\bibfield  {journal} {\bibinfo
   {journal} {Phys. Rev. D}\ }\textbf {\bibinfo {volume} {104}},\ \bibinfo
  {pages} {114016} (\bibinfo {year} {2021})},\ \Eprint
  {https://arxiv.org/abs/2107.05352} {arXiv:2107.05352 [hep-ph]} \BibitemShut
  {NoStop}%
\bibitem [{\citenamefont {Horak}\ \emph {et~al.}(2022)\citenamefont {Horak},
  \citenamefont {Ihssen}, \citenamefont {Papavassiliou}, \citenamefont
  {Pawlowski}, \citenamefont {Weber},\ and\ \citenamefont
  {Wetterich}}]{Horak:2022aqx}%
  \BibitemOpen
  \bibfield  {author} {\bibinfo {author} {\bibfnamefont {J.}~\bibnamefont
  {Horak}}, \bibinfo {author} {\bibfnamefont {F.}~\bibnamefont {Ihssen}},
  \bibinfo {author} {\bibfnamefont {J.}~\bibnamefont {Papavassiliou}}, \bibinfo
  {author} {\bibfnamefont {J.~M.}\ \bibnamefont {Pawlowski}}, \bibinfo {author}
  {\bibfnamefont {A.}~\bibnamefont {Weber}},\ and\ \bibinfo {author}
  {\bibfnamefont {C.}~\bibnamefont {Wetterich}},\ }\bibfield  {title} {\bibinfo
  {title} {{Gluon condensates and effective gluon mass}},\ }\href
  {https://doi.org/10.21468/SciPostPhys.13.2.042} {\bibfield  {journal}
  {\bibinfo  {journal} {SciPost Phys.}\ }\textbf {\bibinfo {volume} {13}},\
  \bibinfo {pages} {042} (\bibinfo {year} {2022})},\ \Eprint
  {https://arxiv.org/abs/2201.09747} {arXiv:2201.09747 [hep-ph]} \BibitemShut
  {NoStop}%
\bibitem [{\citenamefont {Ferreira}\ \emph {et~al.}(2025)\citenamefont
  {Ferreira}, \citenamefont {Papavassiliou}, \citenamefont {Pawlowski},\ and\
  \citenamefont {Wink}}]{Ferreira:2025tzo}%
  \BibitemOpen
  \bibfield  {author} {\bibinfo {author} {\bibfnamefont {M.~N.}\ \bibnamefont
  {Ferreira}}, \bibinfo {author} {\bibfnamefont {J.}~\bibnamefont
  {Papavassiliou}}, \bibinfo {author} {\bibfnamefont {J.~M.}\ \bibnamefont
  {Pawlowski}},\ and\ \bibinfo {author} {\bibfnamefont {N.}~\bibnamefont
  {Wink}},\ }\bibfield  {title} {\bibinfo {title} {{Physics of the gluon mass
  gap}},\ }\href {https://doi.org/10.1140/epjc/s10052-025-15027-7} {\bibfield
  {journal} {\bibinfo  {journal} {Eur. Phys. J. C}\ }\textbf {\bibinfo {volume}
  {85}},\ \bibinfo {pages} {1339} (\bibinfo {year} {2025})},\ \Eprint
  {https://arxiv.org/abs/2508.20568} {arXiv:2508.20568 [hep-ph]} \BibitemShut
  {NoStop}%
\bibitem [{\citenamefont {Junior}\ \emph {et~al.}(2025)\citenamefont {Junior},
  \citenamefont {Krein}, \citenamefont {Oxman},\ and\ \citenamefont
  {Soares}}]{Junior:2025gxg}%
  \BibitemOpen
  \bibfield  {author} {\bibinfo {author} {\bibfnamefont {D.~R.}\ \bibnamefont
  {Junior}}, \bibinfo {author} {\bibfnamefont {G.}~\bibnamefont {Krein}},
  \bibinfo {author} {\bibfnamefont {L.~E.}\ \bibnamefont {Oxman}},\ and\
  \bibinfo {author} {\bibfnamefont {B.~R.}\ \bibnamefont {Soares}},\ }\bibfield
   {title} {\bibinfo {title} {{Center vortices and the emergence of a gluon
  mass scale}},\ }\href@noop {} {\  (\bibinfo {year} {2025})},\ \Eprint
  {https://arxiv.org/abs/2510.19103} {arXiv:2510.19103 [hep-th]} \BibitemShut
  {NoStop}%
\bibitem [{\citenamefont {Tissier}\ and\ \citenamefont
  {Wschebor}(2010)}]{Tissier:2010ts}%
  \BibitemOpen
  \bibfield  {author} {\bibinfo {author} {\bibfnamefont {M.}~\bibnamefont
  {Tissier}}\ and\ \bibinfo {author} {\bibfnamefont {N.}~\bibnamefont
  {Wschebor}},\ }\bibfield  {title} {\bibinfo {title} {{Infrared propagators of
  Yang-Mills theory from perturbation theory}},\ }\href
  {https://doi.org/10.1103/PhysRevD.82.101701} {\bibfield  {journal} {\bibinfo
  {journal} {Phys. Rev. D}\ }\textbf {\bibinfo {volume} {82}},\ \bibinfo
  {pages} {101701} (\bibinfo {year} {2010})},\ \Eprint
  {https://arxiv.org/abs/1004.1607} {arXiv:1004.1607 [hep-ph]} \BibitemShut
  {NoStop}%
\bibitem [{\citenamefont {Tissier}\ and\ \citenamefont
  {Wschebor}(2011)}]{Tissier:2011ey}%
  \BibitemOpen
  \bibfield  {author} {\bibinfo {author} {\bibfnamefont {M.}~\bibnamefont
  {Tissier}}\ and\ \bibinfo {author} {\bibfnamefont {N.}~\bibnamefont
  {Wschebor}},\ }\bibfield  {title} {\bibinfo {title} {{An Infrared Safe
  perturbative approach to Yang-Mills correlators}},\ }\href
  {https://doi.org/10.1103/PhysRevD.84.045018} {\bibfield  {journal} {\bibinfo
  {journal} {Phys. Rev. D}\ }\textbf {\bibinfo {volume} {84}},\ \bibinfo
  {pages} {045018} (\bibinfo {year} {2011})},\ \Eprint
  {https://arxiv.org/abs/1105.2475} {arXiv:1105.2475 [hep-th]} \BibitemShut
  {NoStop}%
\bibitem [{\citenamefont {Gracey}\ \emph {et~al.}(2019)\citenamefont {Gracey},
  \citenamefont {Pel\'aez}, \citenamefont {Reinosa},\ and\ \citenamefont
  {Tissier}}]{Gracey:2019xom}%
  \BibitemOpen
  \bibfield  {author} {\bibinfo {author} {\bibfnamefont {J.~A.}\ \bibnamefont
  {Gracey}}, \bibinfo {author} {\bibfnamefont {M.}~\bibnamefont {Pel\'aez}},
  \bibinfo {author} {\bibfnamefont {U.}~\bibnamefont {Reinosa}},\ and\ \bibinfo
  {author} {\bibfnamefont {M.}~\bibnamefont {Tissier}},\ }\bibfield  {title}
  {\bibinfo {title} {{Two loop calculation of Yang-Mills propagators in the
  Curci-Ferrari model}},\ }\href {https://doi.org/10.1103/PhysRevD.100.034023}
  {\bibfield  {journal} {\bibinfo  {journal} {Phys. Rev. D}\ }\textbf {\bibinfo
  {volume} {100}},\ \bibinfo {pages} {034023} (\bibinfo {year} {2019})},\
  \Eprint {https://arxiv.org/abs/1905.07262} {arXiv:1905.07262 [hep-th]}
  \BibitemShut {NoStop}%
\bibitem [{\citenamefont {Reinosa}(2024)}]{Reinosa:2024vph}%
  \BibitemOpen
  \bibfield  {author} {\bibinfo {author} {\bibfnamefont {U.}~\bibnamefont
  {Reinosa}},\ }\bibfield  {title} {\bibinfo {title} {{Three lectures on the
  Curci-Ferrari model}},\ }in\ \href@noop {} {\emph {\bibinfo {booktitle} {{3rd
  Joint ICTP-SAIFR/ICTP-Trieste Summer School on Particle Physics}}}}\
  (\bibinfo {year} {2024})\ \Eprint {https://arxiv.org/abs/2403.10702}
  {arXiv:2403.10702 [hep-ph]} \BibitemShut {NoStop}%
\bibitem [{\citenamefont {Pel\'aez}\ \emph {et~al.}(2014)\citenamefont
  {Pel\'aez}, \citenamefont {Tissier},\ and\ \citenamefont
  {Wschebor}}]{Pelaez:2014mxa}%
  \BibitemOpen
  \bibfield  {author} {\bibinfo {author} {\bibfnamefont {M.}~\bibnamefont
  {Pel\'aez}}, \bibinfo {author} {\bibfnamefont {M.}~\bibnamefont {Tissier}},\
  and\ \bibinfo {author} {\bibfnamefont {N.}~\bibnamefont {Wschebor}},\
  }\bibfield  {title} {\bibinfo {title} {{Two-point correlation functions of
  QCD in the Landau gauge}},\ }\href
  {https://doi.org/10.1103/PhysRevD.90.065031} {\bibfield  {journal} {\bibinfo
  {journal} {Phys. Rev. D}\ }\textbf {\bibinfo {volume} {90}},\ \bibinfo
  {pages} {065031} (\bibinfo {year} {2014})},\ \Eprint
  {https://arxiv.org/abs/1407.2005} {arXiv:1407.2005 [hep-th]} \BibitemShut
  {NoStop}%
\bibitem [{\citenamefont {Pelaez}\ \emph {et~al.}(2013)\citenamefont {Pelaez},
  \citenamefont {Tissier},\ and\ \citenamefont {Wschebor}}]{Pelaez:2013cpa}%
  \BibitemOpen
  \bibfield  {author} {\bibinfo {author} {\bibfnamefont {M.}~\bibnamefont
  {Pelaez}}, \bibinfo {author} {\bibfnamefont {M.}~\bibnamefont {Tissier}},\
  and\ \bibinfo {author} {\bibfnamefont {N.}~\bibnamefont {Wschebor}},\
  }\bibfield  {title} {\bibinfo {title} {{Three-point correlation functions in
  Yang-Mills theory}},\ }\href {https://doi.org/10.1103/PhysRevD.88.125003}
  {\bibfield  {journal} {\bibinfo  {journal} {Phys. Rev. D}\ }\textbf {\bibinfo
  {volume} {88}},\ \bibinfo {pages} {125003} (\bibinfo {year} {2013})},\
  \Eprint {https://arxiv.org/abs/1310.2594} {arXiv:1310.2594 [hep-th]}
  \BibitemShut {NoStop}%
\bibitem [{\citenamefont {Figueroa}\ and\ \citenamefont
  {Pel{\'a}ez}(2022)}]{Figueroa:2021sjm}%
  \BibitemOpen
  \bibfield  {author} {\bibinfo {author} {\bibfnamefont {F.}~\bibnamefont
  {Figueroa}}\ and\ \bibinfo {author} {\bibfnamefont {M.}~\bibnamefont
  {Pel{\'a}ez}},\ }\bibfield  {title} {\bibinfo {title} {{One-loop unquenched
  three-gluon and ghost-gluon vertices in the Curci-Ferrari model}},\ }\href
  {https://doi.org/10.1103/PhysRevD.105.094005} {\bibfield  {journal} {\bibinfo
   {journal} {Phys. Rev. D}\ }\textbf {\bibinfo {volume} {105}},\ \bibinfo
  {pages} {094005} (\bibinfo {year} {2022})},\ \Eprint
  {https://arxiv.org/abs/2110.09561} {arXiv:2110.09561 [hep-th]} \BibitemShut
  {NoStop}%
\bibitem [{\citenamefont {Pel{\'a}ez}\ \emph {et~al.}(2015)\citenamefont
  {Pel{\'a}ez}, \citenamefont {Tissier},\ and\ \citenamefont
  {Wschebor}}]{Pelaez:2015tba}%
  \BibitemOpen
  \bibfield  {author} {\bibinfo {author} {\bibfnamefont {M.}~\bibnamefont
  {Pel{\'a}ez}}, \bibinfo {author} {\bibfnamefont {M.}~\bibnamefont
  {Tissier}},\ and\ \bibinfo {author} {\bibfnamefont {N.}~\bibnamefont
  {Wschebor}},\ }\bibfield  {title} {\bibinfo {title} {{Quark-gluon vertex from
  the Landau gauge Curci-Ferrari model}},\ }\href
  {https://doi.org/10.1103/PhysRevD.92.045012} {\bibfield  {journal} {\bibinfo
  {journal} {Phys. Rev. D}\ }\textbf {\bibinfo {volume} {92}},\ \bibinfo
  {pages} {045012} (\bibinfo {year} {2015})},\ \Eprint
  {https://arxiv.org/abs/1504.05157} {arXiv:1504.05157 [hep-th]} \BibitemShut
  {NoStop}%
\bibitem [{\citenamefont {Barrios}\ \emph {et~al.}(2020)\citenamefont
  {Barrios}, \citenamefont {Pel\'aez}, \citenamefont {Reinosa},\ and\
  \citenamefont {Wschebor}}]{Barrios:2020ubx}%
  \BibitemOpen
  \bibfield  {author} {\bibinfo {author} {\bibfnamefont {N.}~\bibnamefont
  {Barrios}}, \bibinfo {author} {\bibfnamefont {M.}~\bibnamefont {Pel\'aez}},
  \bibinfo {author} {\bibfnamefont {U.}~\bibnamefont {Reinosa}},\ and\ \bibinfo
  {author} {\bibfnamefont {N.}~\bibnamefont {Wschebor}},\ }\bibfield  {title}
  {\bibinfo {title} {{The ghost-antighost-gluon vertex from the Curci-Ferrari
  model: Two-loop corrections}},\ }\href
  {https://doi.org/10.1103/PhysRevD.102.114016} {\bibfield  {journal} {\bibinfo
   {journal} {Phys. Rev. D}\ }\textbf {\bibinfo {volume} {102}},\ \bibinfo
  {pages} {114016} (\bibinfo {year} {2020})},\ \Eprint
  {https://arxiv.org/abs/2009.00875} {arXiv:2009.00875 [hep-th]} \BibitemShut
  {NoStop}%
\bibitem [{\citenamefont {Barrios}\ \emph {et~al.}(2024)\citenamefont
  {Barrios}, \citenamefont {De~Fabritiis},\ and\ \citenamefont
  {Pel\'aez}}]{Barrios:2024ixj}%
  \BibitemOpen
  \bibfield  {author} {\bibinfo {author} {\bibfnamefont {N.}~\bibnamefont
  {Barrios}}, \bibinfo {author} {\bibfnamefont {P.}~\bibnamefont
  {De~Fabritiis}},\ and\ \bibinfo {author} {\bibfnamefont {M.}~\bibnamefont
  {Pel\'aez}},\ }\bibfield  {title} {\bibinfo {title} {{Four-gluon vertex from
  the Curci-Ferrari model at one-loop order}},\ }\href
  {https://doi.org/10.1103/PhysRevD.109.L091502} {\bibfield  {journal}
  {\bibinfo  {journal} {Phys. Rev. D}\ }\textbf {\bibinfo {volume} {109}},\
  \bibinfo {pages} {L091502} (\bibinfo {year} {2024})},\ \Eprint
  {https://arxiv.org/abs/2403.17056} {arXiv:2403.17056 [hep-th]} \BibitemShut
  {NoStop}%
\bibitem [{\citenamefont {Bopsin}\ \emph {et~al.}(2025)\citenamefont {Bopsin},
  \citenamefont {El-Bennich}, \citenamefont {Krein}, \citenamefont {Serna},\
  and\ \citenamefont {da~Silveira}}]{Bopsin:2025vhz}%
  \BibitemOpen
  \bibfield  {author} {\bibinfo {author} {\bibfnamefont {G.~B.}\ \bibnamefont
  {Bopsin}}, \bibinfo {author} {\bibfnamefont {B.}~\bibnamefont {El-Bennich}},
  \bibinfo {author} {\bibfnamefont {G.}~\bibnamefont {Krein}}, \bibinfo
  {author} {\bibfnamefont {F.~E.}\ \bibnamefont {Serna}},\ and\ \bibinfo
  {author} {\bibfnamefont {R.~C.}\ \bibnamefont {da~Silveira}},\ }\bibfield
  {title} {\bibinfo {title} {{Parton distribution and fragmentation functions
  with massive gluons}},\ }\href {https://doi.org/10.1103/ygrj-pglg} {\bibfield
   {journal} {\bibinfo  {journal} {Phys. Rev. D}\ }\textbf {\bibinfo {volume}
  {112}},\ \bibinfo {pages} {114023} (\bibinfo {year} {2025})},\ \Eprint
  {https://arxiv.org/abs/2507.12544} {arXiv:2507.12544 [hep-ph]} \BibitemShut
  {NoStop}%
\bibitem [{\citenamefont {Alvez}\ \emph {et~al.}(2025)\citenamefont {Alvez},
  \citenamefont {Barrios}, \citenamefont {Ben{\'\i}tez},\ and\ \citenamefont
  {Pel{\'a}ez}}]{Alvez:2025wek}%
  \BibitemOpen
  \bibfield  {author} {\bibinfo {author} {\bibfnamefont {A.}~\bibnamefont
  {Alvez}}, \bibinfo {author} {\bibfnamefont {N.}~\bibnamefont {Barrios}},
  \bibinfo {author} {\bibfnamefont {F.}~\bibnamefont {Ben{\'\i}tez}},\ and\
  \bibinfo {author} {\bibfnamefont {M.}~\bibnamefont {Pel{\'a}ez}},\ }\bibfield
   {title} {\bibinfo {title} {{Nonrelativistic meson masses from the
  Curci-Ferrari model}},\ }\href@noop {} {\  (\bibinfo {year} {2025})},\
  \Eprint {https://arxiv.org/abs/2509.04365} {arXiv:2509.04365 [hep-ph]}
  \BibitemShut {NoStop}%
\bibitem [{\citenamefont {Reinosa}\ \emph {et~al.}(2015)\citenamefont
  {Reinosa}, \citenamefont {Serreau}, \citenamefont {Tissier},\ and\
  \citenamefont {Wschebor}}]{Reinosa:2014ooa}%
  \BibitemOpen
  \bibfield  {author} {\bibinfo {author} {\bibfnamefont {U.}~\bibnamefont
  {Reinosa}}, \bibinfo {author} {\bibfnamefont {J.}~\bibnamefont {Serreau}},
  \bibinfo {author} {\bibfnamefont {M.}~\bibnamefont {Tissier}},\ and\ \bibinfo
  {author} {\bibfnamefont {N.}~\bibnamefont {Wschebor}},\ }\bibfield  {title}
  {\bibinfo {title} {{Deconfinement transition in SU($N$) theories from
  perturbation theory}},\ }\href
  {https://doi.org/10.1016/j.physletb.2015.01.006} {\bibfield  {journal}
  {\bibinfo  {journal} {Phys. Lett. B}\ }\textbf {\bibinfo {volume} {742}},\
  \bibinfo {pages} {61} (\bibinfo {year} {2015})},\ \Eprint
  {https://arxiv.org/abs/1407.6469} {arXiv:1407.6469 [hep-ph]} \BibitemShut
  {NoStop}%
\bibitem [{\citenamefont {van Egmond}\ \emph {et~al.}(2022)\citenamefont {van
  Egmond}, \citenamefont {Reinosa}, \citenamefont {Serreau},\ and\
  \citenamefont {Tissier}}]{vanEgmond:2021jyx}%
  \BibitemOpen
  \bibfield  {author} {\bibinfo {author} {\bibfnamefont {D.~M.}\ \bibnamefont
  {van Egmond}}, \bibinfo {author} {\bibfnamefont {U.}~\bibnamefont {Reinosa}},
  \bibinfo {author} {\bibfnamefont {J.}~\bibnamefont {Serreau}},\ and\ \bibinfo
  {author} {\bibfnamefont {M.}~\bibnamefont {Tissier}},\ }\bibfield  {title}
  {\bibinfo {title} {{A novel background field approach to the
  confinement-deconfinement transition}},\ }\href
  {https://doi.org/10.21468/SciPostPhys.12.3.087} {\bibfield  {journal}
  {\bibinfo  {journal} {SciPost Phys.}\ }\textbf {\bibinfo {volume} {12}},\
  \bibinfo {pages} {087} (\bibinfo {year} {2022})},\ \Eprint
  {https://arxiv.org/abs/2104.08974} {arXiv:2104.08974 [hep-ph]} \BibitemShut
  {NoStop}%
\bibitem [{\citenamefont {van Egmond}\ \emph {et~al.}(2025)\citenamefont {van
  Egmond}, \citenamefont {Oliveira}, \citenamefont {Reinosa}, \citenamefont
  {Serreau}, \citenamefont {Silva},\ and\ \citenamefont
  {Tissier}}]{vanEgmond:2025zxf}%
  \BibitemOpen
  \bibfield  {author} {\bibinfo {author} {\bibfnamefont {D.~M.}\ \bibnamefont
  {van Egmond}}, \bibinfo {author} {\bibfnamefont {O.}~\bibnamefont
  {Oliveira}}, \bibinfo {author} {\bibfnamefont {U.}~\bibnamefont {Reinosa}},
  \bibinfo {author} {\bibfnamefont {J.}~\bibnamefont {Serreau}}, \bibinfo
  {author} {\bibfnamefont {P.~J.}\ \bibnamefont {Silva}},\ and\ \bibinfo
  {author} {\bibfnamefont {M.}~\bibnamefont {Tissier}},\ }\bibfield  {title}
  {\bibinfo {title} {{Center-symmetric Landau gauge, the deconfinement
  transition and the gluon propagator as seen in lattice QCD}},\ }\href
  {https://doi.org/10.22323/1.483.0169} {\bibfield  {journal} {\bibinfo
  {journal} {PoS}\ }\textbf {\bibinfo {volume} {QCHSC24}},\ \bibinfo {pages}
  {169} (\bibinfo {year} {2025})},\ \Eprint {https://arxiv.org/abs/2505.16940}
  {arXiv:2505.16940 [hep-lat]} \BibitemShut {NoStop}%
\bibitem [{\citenamefont {Becchi}\ \emph {et~al.}(1976)\citenamefont {Becchi},
  \citenamefont {Rouet},\ and\ \citenamefont {Stora}}]{Becchi:1975nq}%
  \BibitemOpen
  \bibfield  {author} {\bibinfo {author} {\bibfnamefont {C.}~\bibnamefont
  {Becchi}}, \bibinfo {author} {\bibfnamefont {A.}~\bibnamefont {Rouet}},\ and\
  \bibinfo {author} {\bibfnamefont {R.}~\bibnamefont {Stora}},\ }\bibfield
  {title} {\bibinfo {title} {{Renormalization of Gauge Theories}},\ }\href
  {https://doi.org/10.1016/0003-4916(76)90156-1} {\bibfield  {journal}
  {\bibinfo  {journal} {Annals Phys.}\ }\textbf {\bibinfo {volume} {98}},\
  \bibinfo {pages} {287} (\bibinfo {year} {1976})}\BibitemShut {NoStop}%
\bibitem [{\citenamefont {Tyutin}(1975)}]{Tyutin:1975qk}%
  \BibitemOpen
  \bibfield  {author} {\bibinfo {author} {\bibfnamefont {I.~V.}\ \bibnamefont
  {Tyutin}},\ }\bibfield  {title} {\bibinfo {title} {{Gauge Invariance in Field
  Theory and Statistical Physics in Operator Formalism}},\ }\href@noop {} {\
  (\bibinfo {year} {1975})},\ \Eprint {https://arxiv.org/abs/0812.0580}
  {arXiv:0812.0580 [hep-th]} \BibitemShut {NoStop}%
\bibitem [{\citenamefont {Kugo}\ and\ \citenamefont
  {Ojima}(1979)}]{Kugo:1979gm}%
  \BibitemOpen
  \bibfield  {author} {\bibinfo {author} {\bibfnamefont {T.}~\bibnamefont
  {Kugo}}\ and\ \bibinfo {author} {\bibfnamefont {I.}~\bibnamefont {Ojima}},\
  }\bibfield  {title} {\bibinfo {title} {Local covariant operator formalism of
  nonabelian gauge theories},\ }\href {https://doi.org/10.1143/PTPS.66.1}
  {\bibfield  {journal} {\bibinfo  {journal} {Prog. Theor. Phys. Suppl.}\
  }\textbf {\bibinfo {volume} {66}},\ \bibinfo {pages} {1} (\bibinfo {year}
  {1979})}\BibitemShut {NoStop}%
\bibitem [{\citenamefont {Comitini}\ \emph {et~al.}(2024)\citenamefont
  {Comitini}, \citenamefont {De~Meerleer}, \citenamefont {Dudal},\ and\
  \citenamefont {Sorella}}]{Comitini:2023urc}%
  \BibitemOpen
  \bibfield  {author} {\bibinfo {author} {\bibfnamefont {G.}~\bibnamefont
  {Comitini}}, \bibinfo {author} {\bibfnamefont {T.}~\bibnamefont
  {De~Meerleer}}, \bibinfo {author} {\bibfnamefont {D.}~\bibnamefont {Dudal}},\
  and\ \bibinfo {author} {\bibfnamefont {S.~P.}\ \bibnamefont {Sorella}},\
  }\bibfield  {title} {\bibinfo {title} {{Dynamically massive linear covariant
  gauges: Setup and first results}},\ }\href
  {https://doi.org/10.1103/PhysRevD.109.014037} {\bibfield  {journal} {\bibinfo
   {journal} {Phys. Rev. D}\ }\textbf {\bibinfo {volume} {109}},\ \bibinfo
  {pages} {014037} (\bibinfo {year} {2024})},\ \Eprint
  {https://arxiv.org/abs/2312.07608} {arXiv:2312.07608 [hep-th]} \BibitemShut
  {NoStop}%
\bibitem [{\citenamefont {Cabrera}\ \emph {et~al.}(2026)\citenamefont
  {Cabrera}, \citenamefont {Pel{\'a}ez},\ and\ \citenamefont
  {Tissier}}]{Cabrera:2026arc}%
  \BibitemOpen
  \bibfield  {author} {\bibinfo {author} {\bibfnamefont {S.}~\bibnamefont
  {Cabrera}}, \bibinfo {author} {\bibfnamefont {M.}~\bibnamefont
  {Pel{\'a}ez}},\ and\ \bibinfo {author} {\bibfnamefont {M.}~\bibnamefont
  {Tissier}},\ }\bibfield  {title} {\bibinfo {title} {{Effects of fermions in
  one-loop propagators in the Curci-Ferrari-Delbourgo-Jarvis gauge}},\
  }\href@noop {} {\  (\bibinfo {year} {2026})},\ \Eprint
  {https://arxiv.org/abs/2603.08460} {arXiv:2603.08460 [hep-th]} \BibitemShut
  {NoStop}%
\bibitem [{\citenamefont {Amemiya}\ and\ \citenamefont
  {Suganuma}(1999)}]{Amemiya:1998jz}%
  \BibitemOpen
  \bibfield  {author} {\bibinfo {author} {\bibfnamefont {K.}~\bibnamefont
  {Amemiya}}\ and\ \bibinfo {author} {\bibfnamefont {H.}~\bibnamefont
  {Suganuma}},\ }\bibfield  {title} {\bibinfo {title} {{Off diagonal gluon mass
  generation and infrared Abelian dominance in the maximally Abelian gauge in
  lattice QCD}},\ }\href {https://doi.org/10.1103/PhysRevD.60.114509}
  {\bibfield  {journal} {\bibinfo  {journal} {Phys. Rev. D}\ }\textbf {\bibinfo
  {volume} {60}},\ \bibinfo {pages} {114509} (\bibinfo {year} {1999})},\
  \Eprint {https://arxiv.org/abs/hep-lat/9811035} {arXiv:hep-lat/9811035}
  \BibitemShut {NoStop}%
\bibitem [{\citenamefont {Bornyakov}\ \emph
  {et~al.}(2003{\natexlab{a}})\citenamefont {Bornyakov}, \citenamefont
  {Morozov},\ and\ \citenamefont {Polikarpov}}]{Bornyakov:2002vv}%
  \BibitemOpen
  \bibfield  {author} {\bibinfo {author} {\bibfnamefont {V.~G.}\ \bibnamefont
  {Bornyakov}}, \bibinfo {author} {\bibfnamefont {S.~M.}\ \bibnamefont
  {Morozov}},\ and\ \bibinfo {author} {\bibfnamefont {M.~I.}\ \bibnamefont
  {Polikarpov}},\ }\bibfield  {title} {\bibinfo {title} {{Gluon propagators in
  maximal Abelian gauge of SU(2) lattice gauge theory}},\ }\href
  {https://doi.org/10.1016/S0920-5632(03)01648-7} {\bibfield  {journal}
  {\bibinfo  {journal} {Nucl. Phys. B Proc. Suppl.}\ }\textbf {\bibinfo
  {volume} {119}},\ \bibinfo {pages} {733} (\bibinfo {year}
  {2003}{\natexlab{a}})},\ \Eprint {https://arxiv.org/abs/hep-lat/0209031}
  {arXiv:hep-lat/0209031} \BibitemShut {NoStop}%
\bibitem [{\citenamefont {Bornyakov}\ \emph
  {et~al.}(2003{\natexlab{b}})\citenamefont {Bornyakov}, \citenamefont
  {Chernodub}, \citenamefont {Gubarev}, \citenamefont {Morozov},\ and\
  \citenamefont {Polikarpov}}]{Bornyakov:2003ee}%
  \BibitemOpen
  \bibfield  {author} {\bibinfo {author} {\bibfnamefont {V.~G.}\ \bibnamefont
  {Bornyakov}}, \bibinfo {author} {\bibfnamefont {M.~N.}\ \bibnamefont
  {Chernodub}}, \bibinfo {author} {\bibfnamefont {F.~V.}\ \bibnamefont
  {Gubarev}}, \bibinfo {author} {\bibfnamefont {S.~M.}\ \bibnamefont
  {Morozov}},\ and\ \bibinfo {author} {\bibfnamefont {M.~I.}\ \bibnamefont
  {Polikarpov}},\ }\bibfield  {title} {\bibinfo {title} {{Abelian dominance and
  gluon propagators in the maximally Abelian gauge of SU(2) lattice gauge
  theory}},\ }\href {https://doi.org/10.1016/S0370-2693(03)00368-X} {\bibfield
  {journal} {\bibinfo  {journal} {Phys. Lett. B}\ }\textbf {\bibinfo {volume}
  {559}},\ \bibinfo {pages} {214} (\bibinfo {year} {2003}{\natexlab{b}})},\
  \Eprint {https://arxiv.org/abs/hep-lat/0302002} {arXiv:hep-lat/0302002}
  \BibitemShut {NoStop}%
\bibitem [{\citenamefont {Huber}\ \emph {et~al.}(2010)\citenamefont {Huber},
  \citenamefont {Schwenzer},\ and\ \citenamefont {Alkofer}}]{Huber:2010tvj}%
  \BibitemOpen
  \bibfield  {author} {\bibinfo {author} {\bibfnamefont {M.~Q.}\ \bibnamefont
  {Huber}}, \bibinfo {author} {\bibfnamefont {K.}~\bibnamefont {Schwenzer}},\
  and\ \bibinfo {author} {\bibfnamefont {R.}~\bibnamefont {Alkofer}},\
  }\bibfield  {title} {\bibinfo {title} {{On the infrared scaling solution of
  SU(N) Yang-Mills theories in the maximally Abelian gauge}},\ }\href
  {https://doi.org/10.1140/epjc/s10052-010-1371-x} {\bibfield  {journal}
  {\bibinfo  {journal} {Eur. Phys. J. C}\ }\textbf {\bibinfo {volume} {68}},\
  \bibinfo {pages} {581} (\bibinfo {year} {2010})},\ \Eprint
  {https://arxiv.org/abs/0904.1873} {arXiv:0904.1873 [hep-th]} \BibitemShut
  {NoStop}%
\bibitem [{\citenamefont {Gongyo}\ \emph {et~al.}(2012)\citenamefont {Gongyo},
  \citenamefont {Iritani},\ and\ \citenamefont {Suganuma}}]{Gongyo:2012jb}%
  \BibitemOpen
  \bibfield  {author} {\bibinfo {author} {\bibfnamefont {S.}~\bibnamefont
  {Gongyo}}, \bibinfo {author} {\bibfnamefont {T.}~\bibnamefont {Iritani}},\
  and\ \bibinfo {author} {\bibfnamefont {H.}~\bibnamefont {Suganuma}},\
  }\bibfield  {title} {\bibinfo {title} {{Off-diagonal Gluon Mass Generation
  and Infrared Abelian Dominance in Maximally Abelian Gauge in SU(3) Lattice
  QCD}},\ }\href {https://doi.org/10.1103/PhysRevD.86.094018} {\bibfield
  {journal} {\bibinfo  {journal} {Phys. Rev. D}\ }\textbf {\bibinfo {volume}
  {86}},\ \bibinfo {pages} {094018} (\bibinfo {year} {2012})},\ \Eprint
  {https://arxiv.org/abs/1207.4377} {arXiv:1207.4377 [hep-lat]} \BibitemShut
  {NoStop}%
\bibitem [{\citenamefont {Kondo}\ and\ \citenamefont
  {Shinohara}(2001)}]{Kondo:2000zva}%
  \BibitemOpen
  \bibfield  {author} {\bibinfo {author} {\bibfnamefont {K.-I.}\ \bibnamefont
  {Kondo}}\ and\ \bibinfo {author} {\bibfnamefont {T.}~\bibnamefont
  {Shinohara}},\ }\bibfield  {title} {\bibinfo {title} {{Renormalizable Abelian
  projected effective gauge theory derived from quantum chromodynamics}},\
  }\href {https://doi.org/10.1143/PTP.105.649} {\bibfield  {journal} {\bibinfo
  {journal} {Prog. Theor. Phys.}\ }\textbf {\bibinfo {volume} {105}},\ \bibinfo
  {pages} {649} (\bibinfo {year} {2001})},\ \Eprint
  {https://arxiv.org/abs/hep-th/0005125} {arXiv:hep-th/0005125} \BibitemShut
  {NoStop}%
\bibitem [{\citenamefont {Fazio}\ \emph {et~al.}(2001)\citenamefont {Fazio},
  \citenamefont {Lemes}, \citenamefont {Sarandy},\ and\ \citenamefont
  {Sorella}}]{Fazio:2001rm}%
  \BibitemOpen
  \bibfield  {author} {\bibinfo {author} {\bibfnamefont {A.~R.}\ \bibnamefont
  {Fazio}}, \bibinfo {author} {\bibfnamefont {V.~E.~R.}\ \bibnamefont {Lemes}},
  \bibinfo {author} {\bibfnamefont {M.~S.}\ \bibnamefont {Sarandy}},\ and\
  \bibinfo {author} {\bibfnamefont {S.~P.}\ \bibnamefont {Sorella}},\
  }\bibfield  {title} {\bibinfo {title} {{The Diagonal ghost equation Ward
  identity for Yang-Mills theories in the maximal Abelian gauge}},\ }\href
  {https://doi.org/10.1103/PhysRevD.64.085003} {\bibfield  {journal} {\bibinfo
  {journal} {Phys. Rev. D}\ }\textbf {\bibinfo {volume} {64}},\ \bibinfo
  {pages} {085003} (\bibinfo {year} {2001})},\ \Eprint
  {https://arxiv.org/abs/hep-th/0105060} {arXiv:hep-th/0105060} \BibitemShut
  {NoStop}%
\bibitem [{\citenamefont {Shinohara}\ \emph {et~al.}(2003)\citenamefont
  {Shinohara}, \citenamefont {Imai},\ and\ \citenamefont
  {Kondo}}]{Shinohara:2001cw}%
  \BibitemOpen
  \bibfield  {author} {\bibinfo {author} {\bibfnamefont {T.}~\bibnamefont
  {Shinohara}}, \bibinfo {author} {\bibfnamefont {T.}~\bibnamefont {Imai}},\
  and\ \bibinfo {author} {\bibfnamefont {K.-I.}\ \bibnamefont {Kondo}},\
  }\bibfield  {title} {\bibinfo {title} {{The Most general and renormalizable
  maximal Abelian gauge}},\ }\href {https://doi.org/10.1142/S0217751X03016008}
  {\bibfield  {journal} {\bibinfo  {journal} {Int. J. Mod. Phys. A}\ }\textbf
  {\bibinfo {volume} {18}},\ \bibinfo {pages} {5733} (\bibinfo {year}
  {2003})},\ \Eprint {https://arxiv.org/abs/hep-th/0105268}
  {arXiv:hep-th/0105268} \BibitemShut {NoStop}%
\bibitem [{\citenamefont {Kondo}(2001)}]{Kondo:2001nq}%
  \BibitemOpen
  \bibfield  {author} {\bibinfo {author} {\bibfnamefont {K.-I.}\ \bibnamefont
  {Kondo}},\ }\bibfield  {title} {\bibinfo {title} {{Vacuum condensate of mass
  dimension 2 as the origin of mass gap and quark confinement}},\ }\href
  {https://doi.org/10.1016/S0370-2693(01)00817-6} {\bibfield  {journal}
  {\bibinfo  {journal} {Phys. Lett. B}\ }\textbf {\bibinfo {volume} {514}},\
  \bibinfo {pages} {335} (\bibinfo {year} {2001})},\ \Eprint
  {https://arxiv.org/abs/hep-th/0105299} {arXiv:hep-th/0105299} \BibitemShut
  {NoStop}%
\bibitem [{\citenamefont {Dudal}\ and\ \citenamefont
  {Verschelde}(2003)}]{Dudal:2002xe}%
  \BibitemOpen
  \bibfield  {author} {\bibinfo {author} {\bibfnamefont {D.}~\bibnamefont
  {Dudal}}\ and\ \bibinfo {author} {\bibfnamefont {H.}~\bibnamefont
  {Verschelde}},\ }\bibfield  {title} {\bibinfo {title} {{On ghost
  condensation, mass generation and Abelian dominance in the maximal Abelian
  gauge}},\ }\href {https://doi.org/10.1088/0305-4470/36/31/312} {\bibfield
  {journal} {\bibinfo  {journal} {J. Phys. A}\ }\textbf {\bibinfo {volume}
  {36}},\ \bibinfo {pages} {8507} (\bibinfo {year} {2003})},\ \Eprint
  {https://arxiv.org/abs/hep-th/0209025} {arXiv:hep-th/0209025} \BibitemShut
  {NoStop}%
\bibitem [{\citenamefont {Dudal}\ \emph {et~al.}(2004)\citenamefont {Dudal},
  \citenamefont {Gracey}, \citenamefont {Lemes}, \citenamefont {Sarandy},
  \citenamefont {Sobreiro}, \citenamefont {Sorella},\ and\ \citenamefont
  {Verschelde}}]{Dudal:2004rx}%
  \BibitemOpen
  \bibfield  {author} {\bibinfo {author} {\bibfnamefont {D.}~\bibnamefont
  {Dudal}}, \bibinfo {author} {\bibfnamefont {J.~A.}\ \bibnamefont {Gracey}},
  \bibinfo {author} {\bibfnamefont {V.~E.~R.}\ \bibnamefont {Lemes}}, \bibinfo
  {author} {\bibfnamefont {M.~S.}\ \bibnamefont {Sarandy}}, \bibinfo {author}
  {\bibfnamefont {R.~F.}\ \bibnamefont {Sobreiro}}, \bibinfo {author}
  {\bibfnamefont {S.~P.}\ \bibnamefont {Sorella}},\ and\ \bibinfo {author}
  {\bibfnamefont {H.}~\bibnamefont {Verschelde}},\ }\bibfield  {title}
  {\bibinfo {title} {{An Analytic study of the off-diagonal mass generation for
  Yang-Mills theories in the maximal Abelian gauge}},\ }\href
  {https://doi.org/10.1103/PhysRevD.70.114038} {\bibfield  {journal} {\bibinfo
  {journal} {Phys. Rev. D}\ }\textbf {\bibinfo {volume} {70}},\ \bibinfo
  {pages} {114038} (\bibinfo {year} {2004})},\ \Eprint
  {https://arxiv.org/abs/hep-th/0406132} {arXiv:hep-th/0406132} \BibitemShut
  {NoStop}%
\bibitem [{\citenamefont {'t~Hooft}(1981)}]{tHooft:1981bkw}%
  \BibitemOpen
  \bibfield  {author} {\bibinfo {author} {\bibfnamefont {G.}~\bibnamefont
  {'t~Hooft}},\ }\bibfield  {title} {\bibinfo {title} {{Topology of the Gauge
  Condition and New Confinement Phases in Nonabelian Gauge Theories}},\ }\href
  {https://doi.org/10.1016/0550-3213(81)90442-9} {\bibfield  {journal}
  {\bibinfo  {journal} {Nucl. Phys. B}\ }\textbf {\bibinfo {volume} {190}},\
  \bibinfo {pages} {455} (\bibinfo {year} {1981})}\BibitemShut {NoStop}%
\bibitem [{\citenamefont {Ezawa}\ and\ \citenamefont
  {Iwazaki}(1982{\natexlab{b}})}]{Ezawa:1982bf}%
  \BibitemOpen
  \bibfield  {author} {\bibinfo {author} {\bibfnamefont {Z.~F.}\ \bibnamefont
  {Ezawa}}\ and\ \bibinfo {author} {\bibfnamefont {A.}~\bibnamefont
  {Iwazaki}},\ }\bibfield  {title} {\bibinfo {title} {{Abelian Dominance and
  Quark Confinement in Yang-Mills Theories}},\ }\href
  {https://doi.org/10.1103/PhysRevD.25.2681} {\bibfield  {journal} {\bibinfo
  {journal} {Phys. Rev. D}\ }\textbf {\bibinfo {volume} {25}},\ \bibinfo
  {pages} {2681} (\bibinfo {year} {1982}{\natexlab{b}})}\BibitemShut {NoStop}%
\bibitem [{\citenamefont {Suzuki}\ and\ \citenamefont
  {Yotsuyanagi}(1990{\natexlab{b}})}]{Suzuki:1989gp}%
  \BibitemOpen
  \bibfield  {author} {\bibinfo {author} {\bibfnamefont {T.}~\bibnamefont
  {Suzuki}}\ and\ \bibinfo {author} {\bibfnamefont {I.}~\bibnamefont
  {Yotsuyanagi}},\ }\bibfield  {title} {\bibinfo {title} {{A possible evidence
  for Abelian dominance in quark confinement}},\ }\href
  {https://doi.org/10.1103/PhysRevD.42.4257} {\bibfield  {journal} {\bibinfo
  {journal} {Phys. Rev. D}\ }\textbf {\bibinfo {volume} {42}},\ \bibinfo
  {pages} {4257} (\bibinfo {year} {1990}{\natexlab{b}})}\BibitemShut {NoStop}%
\bibitem [{\citenamefont {Kondo}(1998)}]{Kondo:1997pc}%
  \BibitemOpen
  \bibfield  {author} {\bibinfo {author} {\bibfnamefont {K.-I.}\ \bibnamefont
  {Kondo}},\ }\bibfield  {title} {\bibinfo {title} {{Abelian projected
  effective gauge theory of QCD with asymptotic freedom and quark
  confinement}},\ }\href {https://doi.org/10.1103/PhysRevD.57.7467} {\bibfield
  {journal} {\bibinfo  {journal} {Phys. Rev. D}\ }\textbf {\bibinfo {volume}
  {57}},\ \bibinfo {pages} {7467} (\bibinfo {year} {1998})},\ \Eprint
  {https://arxiv.org/abs/hep-th/9709109} {arXiv:hep-th/9709109} \BibitemShut
  {NoStop}%
\bibitem [{\citenamefont {Capri}\ \emph {et~al.}(2006)\citenamefont {Capri},
  \citenamefont {Lemes}, \citenamefont {Sobreiro}, \citenamefont {Sorella},\
  and\ \citenamefont {Thibes}}]{Capri:2006cz}%
  \BibitemOpen
  \bibfield  {author} {\bibinfo {author} {\bibfnamefont {M.~A.~L.}\
  \bibnamefont {Capri}}, \bibinfo {author} {\bibfnamefont {V.~E.~R.}\
  \bibnamefont {Lemes}}, \bibinfo {author} {\bibfnamefont {R.~F.}\ \bibnamefont
  {Sobreiro}}, \bibinfo {author} {\bibfnamefont {S.~P.}\ \bibnamefont
  {Sorella}},\ and\ \bibinfo {author} {\bibfnamefont {R.}~\bibnamefont
  {Thibes}},\ }\bibfield  {title} {\bibinfo {title} {{A Study of the maximal
  Abelian gauge in SU(2) Euclidean Yang-Mills theory in the presence of the
  Gribov horizon}},\ }\href {https://doi.org/10.1103/PhysRevD.74.105007}
  {\bibfield  {journal} {\bibinfo  {journal} {Phys. Rev. D}\ }\textbf {\bibinfo
  {volume} {74}},\ \bibinfo {pages} {105007} (\bibinfo {year} {2006})},\
  \Eprint {https://arxiv.org/abs/hep-th/0609212} {arXiv:hep-th/0609212}
  \BibitemShut {NoStop}%
\bibitem [{\citenamefont {Capri}\ \emph {et~al.}(2010)\citenamefont {Capri},
  \citenamefont {Gomez}, \citenamefont {Guimaraes}, \citenamefont {Lemes},\
  and\ \citenamefont {Sorella}}]{Capri:2010an}%
  \BibitemOpen
  \bibfield  {author} {\bibinfo {author} {\bibfnamefont {M.~A.~L.}\
  \bibnamefont {Capri}}, \bibinfo {author} {\bibfnamefont {A.~J.}\ \bibnamefont
  {Gomez}}, \bibinfo {author} {\bibfnamefont {M.~S.}\ \bibnamefont
  {Guimaraes}}, \bibinfo {author} {\bibfnamefont {V.~E.~R.}\ \bibnamefont
  {Lemes}},\ and\ \bibinfo {author} {\bibfnamefont {S.~P.}\ \bibnamefont
  {Sorella}},\ }\bibfield  {title} {\bibinfo {title} {{Study of the properties
  of the Gribov region in SU(N) Euclidean Yang-Mills theories in the maximal
  Abelian gauge}},\ }\href {https://doi.org/10.1088/1751-8113/43/24/245402}
  {\bibfield  {journal} {\bibinfo  {journal} {J. Phys. A}\ }\textbf {\bibinfo
  {volume} {43}},\ \bibinfo {pages} {245402} (\bibinfo {year} {2010})},\
  \Eprint {https://arxiv.org/abs/1002.1659} {arXiv:1002.1659 [hep-th]}
  \BibitemShut {NoStop}%
\bibitem [{\citenamefont {Capri}\ \emph {et~al.}(2013)\citenamefont {Capri},
  \citenamefont {Dudal}, \citenamefont {Guimaraes}, \citenamefont {Palhares},\
  and\ \citenamefont {Sorella}}]{Capri:2012wx}%
  \BibitemOpen
  \bibfield  {author} {\bibinfo {author} {\bibfnamefont {M.~A.~L.}\
  \bibnamefont {Capri}}, \bibinfo {author} {\bibfnamefont {D.}~\bibnamefont
  {Dudal}}, \bibinfo {author} {\bibfnamefont {M.~S.}\ \bibnamefont
  {Guimaraes}}, \bibinfo {author} {\bibfnamefont {L.~F.}\ \bibnamefont
  {Palhares}},\ and\ \bibinfo {author} {\bibfnamefont {S.~P.}\ \bibnamefont
  {Sorella}},\ }\bibfield  {title} {\bibinfo {title} {{An all-order proof of
  the equivalence between Gribov's no-pole and Zwanziger's horizon
  conditions}},\ }\href {https://doi.org/10.1016/j.physletb.2013.01.039}
  {\bibfield  {journal} {\bibinfo  {journal} {Phys. Lett.}\ }\textbf {\bibinfo
  {volume} {B719}},\ \bibinfo {pages} {448} (\bibinfo {year} {2013})},\ \Eprint
  {https://arxiv.org/abs/1212.2419} {arXiv:1212.2419 [hep-th]} \BibitemShut
  {NoStop}%
%%CITATION = ARXIV:1212.2419;%%
\bibitem [{\citenamefont {Capri}\ \emph
  {et~al.}(2015{\natexlab{b}})\citenamefont {Capri}, \citenamefont
  {Fiorentini},\ and\ \citenamefont {Sorella}}]{Capri:2015pfa}%
  \BibitemOpen
  \bibfield  {author} {\bibinfo {author} {\bibfnamefont {M.~A.~L.}\
  \bibnamefont {Capri}}, \bibinfo {author} {\bibfnamefont {D.}~\bibnamefont
  {Fiorentini}},\ and\ \bibinfo {author} {\bibfnamefont {S.~P.}\ \bibnamefont
  {Sorella}},\ }\bibfield  {title} {\bibinfo {title} {{Gribov horizon and
  non-perturbative BRST symmetry in the maximal Abelian gauge}},\ }\href
  {https://doi.org/10.1016/j.physletb.2015.10.032} {\bibfield  {journal}
  {\bibinfo  {journal} {Phys. Lett. B}\ }\textbf {\bibinfo {volume} {751}},\
  \bibinfo {pages} {262} (\bibinfo {year} {2015}{\natexlab{b}})},\ \Eprint
  {https://arxiv.org/abs/1507.05481} {arXiv:1507.05481 [hep-th]} \BibitemShut
  {NoStop}%
\bibitem [{\citenamefont {Morozov}\ and\ \citenamefont
  {Rogalyov}(2005)}]{Morozov:2005kh}%
  \BibitemOpen
  \bibfield  {author} {\bibinfo {author} {\bibfnamefont {S.~M.}\ \bibnamefont
  {Morozov}}\ and\ \bibinfo {author} {\bibfnamefont {R.~N.}\ \bibnamefont
  {Rogalyov}},\ }\bibfield  {title} {\bibinfo {title} {{Ultraviolet behavior of
  the gluon propagator in the maximal Abelian gauge}},\ }\href
  {https://doi.org/10.1103/PhysRevD.71.085016} {\bibfield  {journal} {\bibinfo
  {journal} {Phys. Rev. D}\ }\textbf {\bibinfo {volume} {71}},\ \bibinfo
  {pages} {085016} (\bibinfo {year} {2005})},\ \Eprint
  {https://arxiv.org/abs/hep-ph/0501122} {arXiv:hep-ph/0501122} \BibitemShut
  {NoStop}%
\bibitem [{\citenamefont {Gongyo}\ and\ \citenamefont
  {Suganuma}(2013)}]{Gongyo:2013sha}%
  \BibitemOpen
  \bibfield  {author} {\bibinfo {author} {\bibfnamefont {S.}~\bibnamefont
  {Gongyo}}\ and\ \bibinfo {author} {\bibfnamefont {H.}~\bibnamefont
  {Suganuma}},\ }\bibfield  {title} {\bibinfo {title} {{Gluon Propagators in
  Maximally Abelian Gauge in SU(3) Lattice QCD}},\ }\href
  {https://doi.org/10.1103/PhysRevD.87.074506} {\bibfield  {journal} {\bibinfo
  {journal} {Phys. Rev. D}\ }\textbf {\bibinfo {volume} {87}},\ \bibinfo
  {pages} {074506} (\bibinfo {year} {2013})},\ \Eprint
  {https://arxiv.org/abs/1302.6181} {arXiv:1302.6181 [hep-lat]} \BibitemShut
  {NoStop}%
\end{thebibliography}%

\end{document}